\documentclass[11pt]{article}
 \pdfoutput=1   
\usepackage{color} 
\usepackage{amsmath,cite,multirow}
\usepackage{tikz}
\usetikzlibrary{arrows, intersections,decorations.markings,decorations.pathmorphing,patterns,shapes.geometric,calc}
\newcommand{\midarrow}{\tikz \draw[-triangle 90] (0,0) -- +(.1,0);}

\DeclareFontFamily{U}{rsf}{} \DeclareFontShape{U}{rsf}{m}{n}{
  <5> <6> rsfs5 <7> <8> <9> rsfs7 <10-> rsfs10}{}
\DeclareMathAlphabet\Scr{U}{rsf}{m}{n} \makeatletter
\@addtoreset{equation}{section} \makeatother

\def\be{\begin{equation}}
\def\ee{\end{equation}}
\def\ba{\begin{array}}
\def\ea{\end{array}}

\newcommand{\bea}{\begin{eqnarray}}
\newcommand{\eea}{\end{eqnarray}}

\usepackage{ifpdf}
\ifx\pdfoutput\undefined
   \pdffalse
\else
   \pdfoutput=1
   \pdftrue
  \usepackage[pdftex]{hyperref}
\newcommand{\chgdone}[2]{#2}

\begin{document}
\title{Lepton non-universality in $B$ decays and fermion mass structure}
\author{B. Grinstein,$^a$ S. Pokorski,$^b$ and G.G.Ross,$^c$}

\maketitle
\begin{center}
\noindent $^a$Department of Physics, UC San Diego, 9500 Gilman Dr,
La Jolla CA, 92093, USA\\
\noindent $^b$Institute of Theoretical Physics, Faculty of Physics, University of Warsaw, 
\break \phantom{m} ul.~Pasteura 5, PL--02--093 Warsaw, Poland\\
\noindent $^c$Rudolf Peierls Centre for Theoretical Physics, Clarendon Laboratory, \break \phantom{m} University of Oxford, Parks Road, Oxford OX1 3PU, UK.
\end{center}
\vspace{-3.5in}
\hfill OUTP-18-06P
\vspace{3.5in}

\begin{abstract}
We consider the possibility that the neutral-current $B$ anomalies are due to radiative corrections generated by Yukawa interactions of quarks and  leptons with new vector-like quark and lepton electroweak doublets and new Standard Model singlet scalars. We show that the restricted interactions needed can result from an underlying Abelian family symmetry and that the same symmetry can give rise to an acceptable pattern of quark and charged lepton masses and mixings, providing a bridge between the non-universality observed in the $B$-sector and that of the fermion mass matrices. We construct two simple models, one with a single singlet scalar in which the flavour changing comes from quark and lepton mixing and one with an additional scalar in which the flavour changing can come from both fermion and scalar mixing. We show that for the case the new quarks are much heavier than the new leptons and scalars the B anomalies can be due to box diagrams with couplings in the perturbative regime consistent with the bounds coming from $B_s-\bar B_s$, $K-\bar K$ and $D-\bar D$ mixing as well as other lepton family number violating processes. The new states can be dark matter candidates and, in the two scalar model with a light scalar of $O(60)\;$GeV and vector-like lepton of $O(100)\;$GeV, there can be a simultaneous explanation of the $B$-anomalies, the muon anomalous magnetic moment and the dark matter abundance.\end{abstract}
\section{Introduction}
Recent measurements of $B$ decays {\cite{Aaij:2014ora,
    Aaij:2017vbb, Aaij:2013qta, Aaij:2015oid, Wehle:2016yoi,
    Aaij:2014pli, Aaij:2015esa}} have indicated there may be
departures from the Standard Model (SM) predictions for processes
induced by the neutral currrent quark-level transition
$b\rightarrow s\mu^+\mu^-$. Indeed an effective field theory analysis
indicates that the inclusion of the operators\footnote{$P_{L,R}$ are
  the left- and right-handed projection operators} 
\be
O_9={\alpha_{EM}\over 4\pi}[\bar
s\gamma^{\nu}P_Lb][\bar\mu\gamma_{\nu}\mu],\;\;\;O_{10}={\alpha_{EM}\over
  4\pi}[\bar s\gamma^{\nu}P_Lb][\bar\mu\gamma_{\nu}\gamma_5\mu]
\label{ops}
\ee
that violate lepton universality can improve the global fit by
$4-5\sigma$ {\cite{Capdevila:2017bsm, Altmannshofer:2017yso,
  DAmico:2017mtc, Hiller:2017bzc, Geng:2017svp, Ciuchini:2017mik}}. 

A departure from lepton universality is not really a surprise as
non-universality in fermion masses and mixings is manifest. Given this
it is relevant to ask whether the non-universal behaviour in B-decays
may be related to the observed fermion mass structure. The most
promising explanation for the origin of fermion masses and mixings is
through a spontaneously broken family symmetry in which the order
parameters for family symmetry breaking are the vacuum expectation
values (vevs) of scalar (familon) fields, $\phi_i$, that transform
non-trivially under the family symmetry. Then a particular entry in
the fermion mass matrix is proportional to some power of the familon
fields vevs, the underlying (higher dimensional) amplitude requiring
the insertion of the familons to be family symmetry invariant. In the
underlying theory the higher dimension amplitude is generated by the
Yukawa couplings of SM fermions to the familons and new heavy
vector-like fermions (the Froggatt-Nielsen mechanism \cite{Froggatt:1978nt}).

How could this be related to non-universality in B-decays? The neutral
current process occurs first in the SM at one-loop order and is CKM
suppressed. As a result the new physics responsible for
non-universality must be small, consistent with a radiative correction
involving new heavy states\cite{Gripaios:2015gra, Cline:2017qqu,Poh:2017tfo,Dhargyal:2017vcu,Dhargyal:2018bbc }. In particular it could be due to virtual
processes involving heavy vectorlike fermions and familons associated
with a family symmetry.  Neutral current corrections involving loop
effects of new scalars and fermions have already been explored in
detail by Arnan, Hofer, Mescia and Crivellin \cite{Arnan:2016cpy} for
a wide variety of SM representations for the scalars and
quarks.\footnote{{Neutral current corrections arising from loop effects
  in R-parity violating SUSY extensions of the SM have been
  investigated in Ref.~\cite{Das:2017kfo}. This approach does not lend
itself to an interpretation as a theory of flavor.}} Their
overall conclusion is that to explain the $b\rightarrow s\mu^+\mu^-$ anomaly the couplings required are forced to be uncomfortably large in comparison with the bounds coming from $B_s-\bar{B}_s$ mixing unless the scalars carry SM
quantum numbers and the heavy fermions are Majorana fermions. This
does not allow for a familon identification of the scalar in which
case there is no obvious relation of the neutral current B-anomalies
to the fermion mass structure. 

In this paper we re-examine in detail the possibility that the
anomalous $B$ neutral current decays can be due to new heavy
vector-like quarks and leptons coupled to Standard ~Model states via
new scalar fields. We show that, allowing for non-degenerate masses of these states, the constraints coming from $B_s-\bar{B}_s$ mixing are relaxed. We also show that the constraint can be further relaxed and indeed eliminated if the possibility of flavour changing effects coming from the scalar sector is included. We investigate what type of family symmetry,
capable of organising the quark and charged lepton masses and mixing,
can generate the required structure without violating the many
stringent constraints on flavour changing neutral currents. As we will
see, this can be done but with a structure that is not normally
considered when building fermion mass models.

Note that there is also an indication of a departure from SM predictions at the
$3.9\sigma$ level \cite{Amhis:2016xyh} in charged current processes
induced by the quark-level transitions $b\rightarrow cl\nu$
\cite{Lees:2012xj, Lees:2013uzd, Huschle:2015rga, Sato:2016svk,
  Hirose:2016wfn, Hirose:2017dxl, Aaij:2015yra , Aaij:2017uff ,
  Aaij:2017deq , Aaij:2017tyk}.  It is remarkable that it is possible
to construct a simultaneous explanation of both charged and neutral
current effects consistent with all present measurements, the most
plausible being due to the exchange of new vector leptoquark
states~\cite{Alonso:2015sja, Hiller:2014yaa, Barbieri:2015yvd, Assad:2017iib,
  DiLuzio:2017vat, Calibbi:2017qbu, Barbieri:2017tuq , Blanke:2018sro
  , ACrivellin:2018qbd, Greljo:2018ogz}. However,
as the charged current processes arise at tree level, such
non-universality cannot be explained by perturbative loop
effects. Thus, in pursuing the familon explanation, we must assume
that the signals for non-universality in charged current processes are
spurious or that some other source of new physics is present.

In Sec.~\ref{intro} we briefly review the structure and calculation
of the box diagrams, involving a heavy vector-like quark and lepton
doublet together with a SM singlet scalar, give rise to
$b\rightarrow s\mu^+\mu^-$ and $B_s-\bar{B}_s$ mixing. Sec.~\ref{family} presents a simple family symmetry that generates the
required coupling structure of the new states together with a
realistic pattern of quark and charged lepton masses and mixing and
the possibility that the new states are dark matter candidates. In
Sec.~\ref{newfc} we show how the addition of a second scalar field
generates additional flavour changing effects coming from the scalar
sector alone and calculate the associated box diagrams.  In Sec.~\ref{pheno} we present a phenomenological analysis of the resulting
$B$ anomalies in both schemes concentrating on the case that the
vector-like leptons and scalars are much lighter than the vector-like
quark. We show that for couplings in the perturbative domain all
neutral current anomalies can be accommodated consistent with the
bounds on $B-\bar B,\;K-\bar K$ and $D-\bar D\;$mixing. For the case {that} the
vector-like lepton and the lightest scalar state are of $O(100)\;\text{GeV}$,
the SM prediction for the anomalous magnetic moment of the muon can
also be brought into agreement with the experimental best fit
value. We also discuss the constraints coming from other neutral
flavour changing processes.  {In some of our flavour models
  there is a $Z_2$ symmetry under which SM fields are neutral; the
  lightest of the new particles that transforms non-trivially is
  therefore stable and a candidate for dark matter.} In Sec.~\ref{DM} we discuss the experimental
direct detection signals for the new particles, the
present experimental limits and the dark matter abundance associated with the
new states. Finally in Sec.~\ref{sum} we present
our summary and conclusions. The Appendices present two simple family symmetry
variants capable of eliminating lepton flavour changing processes completely and allowing the lightest new state to decay so that it is no longer a dark matter candidate. 
\section{Box diagrams}\label{intro}

\begin{figure*}[t]
\begin{center}
\begin{tikzpicture}[scale=0.8]
\coordinate (S) at (7.3,0);
\coordinate (A) at (0,0);
\coordinate (B) at (0,1);
\coordinate (B2) at ($2*(B)$);
\coordinate (C) at (2,0);
\coordinate (D) at ($-2*(B)$);

\begin{scope}[very thick, every node/.style={sloped,allow upside down}]
\draw node[left] {$b_L$} (A) -- node {\midarrow}   ++(B2)  -- node {\midarrow}
++(B2)  -- node {\midarrow} ++ (B2) node[left] {$s_L$}; 
\node[left] at ($(A)+3*(B)$) {$\Psi_Q$};
\draw  ($(C)+6*(B)$) node[right] {$\mu_L$}-- node {\midarrow}   ++(D)  -- node {\midarrow}
++(D)  -- node {\midarrow}
++(D) node[right] {$\mu_L$} ; 
\node[right] at ($(C)+3*(B)$) {$\Psi_\ell$};
\draw[dashed] (B2) -- node {\midarrow}+(C);
\node[below] at ($0.5*(C)+(B2)$) {$\Phi$};
\draw[dashed] ($(C)+2*(B2)$) -- node {\midarrow} +($-1*(C)$);
\node[above] at ($0.5*(C)+2*(B2)$) {$\Phi$};

\node[below] at ($(A)+0.5*(C)$) {(a)};

\coordinate (A) at (S);
\coordinate (B) at (0,1);
\coordinate (B2) at ($2*(B)$);
\coordinate (C) at (2,0);
\coordinate (D) at ($-2*(B)$);
\coordinate (E) at ($(A)+(C)$);

\draw (A)  node[left] {$b_L$} -- node {\midarrow}   ++(B2)  -- node {\midarrow}
++(B2)  -- node {\midarrow} ++ (B2) node[left] {$s_L$}; 
\node[left] at ($(A)+3*(B)$) {$\Psi_Q$};

\draw  ($(E)+6*(B)$) node[right] {$b_L$} -- node {\midarrow}   ++(D)  -- node {\midarrow}
++(D)  -- node {\midarrow} ++(D) node[right] {$s_L$} ; 
\node[right] at ($(E)+3*(B)$) {$\Psi_Q$};

\draw[dashed] ($(A)+(B2)$) -- node {\midarrow} +(C);
\node[below] at ($(A)+0.5*(C)+(B2)$) {$\Phi$};
\draw[dashed] ($(A)+2*(B2)+(C)$) -- node {\midarrow} +($-1*(C)$);
\node[above] at ($(A)+0.5*(C)+2*(B2)$) {$\Phi$};
\node[below] at ($(A)+0.5*(C)$) {(b)};

\end{scope}
\end{tikzpicture}

\end{center}
\vspace{-1cm}
\caption{\it  \small Box-diagrams contributing to (a) $b\to s \mu^+\mu^-$ and (b) $B_s\to \bar B_s$, with couplings given in Eq.~(\ref{coup1}).
\label{Box1} }
\end{figure*}
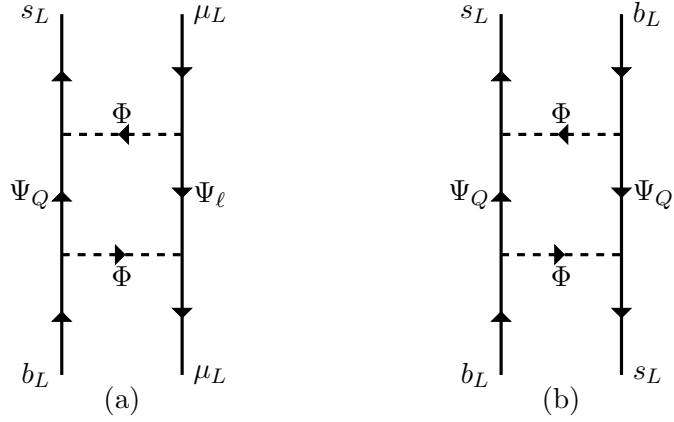

Neutral current $B$ decays generated by scalars can proceed by the generic box diagram shown in Fig.~\ref{Box1}(a). This involves  two additional heavy vector-like quarks and leptons, $\Psi_Q$ and $\Psi_\ell$, and a scalar $\Phi$ with SM quantum numbers given in Table \ref{MS} and interactions described
by the Lagrangian
\be
{\cal L}_{{\mathop{\rm int}} } = \sum_{i=d,s,b} {\Gamma^{m}_i\bar Q^m_i}{P_R}{\Psi_Q}{\Phi} +\sum_{i=e,\mu,\tau} \Gamma^{m,L}_i \bar L^m_i {P_R}{\Psi_\ell}{\Phi} +
{\rm{h}}{\rm{.c.}}
\label{coup1}
\ee
\begin{table}[h]
\centering
\[\begin{array}{*{20}{c}}
&&\vline& {SU(3)}&{SU(2)}&{U(1)}\\
\hline
& {{\Psi _{Q,L}}}&\vline& 3&2&{1/6}\\
& {{\Psi _{Q,R}}}&\vline& 3&2&{1/6}\\
& {{\Psi _{\ell,L}}}&\vline& 1&2&{ - 1/2}\\
& {{\Psi _{\ell,R}}}&\vline& 1&2&{ - 1/2}\\
& \Phi &\vline& 1&1&0
\end{array}\]
\caption{\label{MS}\it \small Multiplet structure of the additional states under $SU(3)\times SU(2)\times U(1)$}
\end{table}
\noindent Here $Q^m_i$ and $L^m_i$ denote the SM left-handed quark and lepton
doublet mass eigenstates with family index $i$.\footnote{{We
    assume the upper components of the doublet $Q^m_i$ are linear
    combinations of the mass eigenstates, $V^{\rm{CKM}\dagger}_{ij}
    U_j$.  }} The  effective Hamiltonian describing the $b\to s\mu^+\mu^-$ transition has the form
\begin{equation}
 {\cal H}_\mathrm{eff}^{\mu^+\mu^-} =   C_9 \mathcal{O}_9+C_{10} 
 \mathcal{O}_{10}
\end{equation}
in terms of the operators given in Eq.~(\ref{ops}). 
The box diagram of Fig.~\ref{Box1}(a) gives the Wilson coefficients
\be
C_9^{{\rm{box}}} = - C_{10}^{{\rm{box}}} = -\frac{\Gamma_{b\to s\mu^+\mu^-}}{{32\pi {\alpha _{{\rm{EM}}}}m_\Phi ^2}}F\left( {x_Q,x_\ell} \right)
\label{mumu}
\ee
where $\chgdone{\Gamma^m_{b\to s\mu^+\mu^-}}{\Gamma_{b\to s\mu^+\mu^-}}={{\Gamma^m _s}\Gamma ^{m*}_b|{\Gamma ^m_\mu }{|^2}}$ and  $x_Q=m_{\Psi_Q}^2/m_\Phi^2$, $x_\ell=m_{\Psi_\ell}^2/m_\Phi^2$, 
 and the dimensionless loop function is given by
\be
F(x,y) = 
\frac{1}{(1-x)(1-y)} + \frac{x^2 \log{x}}{(1-x)^2(x-y)} +\frac{y^2  \log{y}}{(1-y)^2(y-x)} 
\ee

For the  $B_s-\overline{B}_s$ mixing the Hamiltonian has the form
\begin{equation}
{\cal H}_{\rm eff}^{B\bar B}=
 C_{B\bar B}\chgdone{
 ({\bar s}_{\alpha} \gamma^{\mu} P_{L} b_{\alpha})\, ({\bar s}_{\beta} \gamma^{\mu} P_{L} b_{\beta}) }{
 ({\bar s}\gamma^{\mu} P_{L} b)\, ({\bar s}\gamma^{\mu} P_{L} b) }      \,,\\
\end{equation}
where the box diagram,  Fig.~\ref{Box1}(b), gives 
\be
C_{B\bar B} = \frac{\Gamma_{B_s-\overline{B}_s}}{{128{\pi ^2}m_\Phi ^2}}F\left( {x_Q,x_Q} \right)\,,
\label{BBbar}
\ee
with $\Gamma_{B_s-\overline{B}_s}={{{\left( {{\Gamma^m _s}\Gamma ^{m*}_b} \right)}^2}}$.

\section{Family symmetry and flavour changing in the fermion sector}\label{family}
In order to explain the anomaly it is necessary that the box diagram contributes principally to muon pair production, violating lepton universality. Given that the lepton masses also strongly violate lepton universality it is of interest to ask whether the two are related.   
Here we demonstrate that this can be done via a simple  Abelian family symmetry, $U(1)_F$, with  the charge assignments given in Table \ref{MSA}.
\begin{table}[h]
\[
\begin{array}{*{20}{c}}
{}&\vline& {{Q_2}}&{{q_2}}&{{Q_3}}&{{q_3}}&{{L_2}}&{{l_2}}&{{L_3}}&{{l_3}}&\Psi_{Q,(L,R)}&\Psi_{\ell,(L,R)}&\Phi&H&\chi\\
\hline
Q_F&\vline& 2&0&0&0&2&1&-2&-2&2&4&-2&0&1
\end{array}\]
\caption{\label{MSA} \it  \small $U(1)_F$ family symmetry charges for the second and third families}
\end{table}
Here $Q_i$ and $L_i$ denote the left-handed (L) quark and lepton
$SU(2)$ doublets with family index $i$,  $q_i$ and $l_i$ denote the
right-handed  (R) up and down  quark and lepton $SU(2)$ singlets, and
$H$ is the Higgs doublet. We have also added a SM singlet scalar
$\chi$ which acquires a vacuum expectation value (vev) and is responsible for generating the hierarchical structure of the quark and charged lepton masses and mixing.  
\subsection{Heavy vector-like quark and lepton couplings}
With these charge assignments the only renormalisable couplings allowed involving the heavy vector-like states  are given by 
\be
{\cal L}_{{\mathop{\rm int}} } =  {\Gamma_b\bar Q_3}{P_R}{\Psi_Q}{\Phi}  + \Gamma_\mu\bar L_2 {P_R}{\Psi_\ell}\Phi+
{\rm{h}}{\rm{.c.}}
\label{coup2}
\ee
involving only the left-handed quark and lepton ``current" eigenstates. 
The mass eigenstates, $Q^m_i,L^m_i$, in \chgdone{Eq.~(}{Eq.~(}\ref{coup1}) are superpositions of the current eigenstates, the mixing  determined by mixing matrices $V^{Q}$ and $V^{L}$ for the up and down sectors:
\be
Q_i = \begin{pmatrix} V^U_{iu}u+V^U_{ic}c+V^U_{it}t \\[3pt] V^D_{id}d+V^D_{is}s+V^D_{ib}b\end{pmatrix},\qquad L_i = \begin{pmatrix} V^N_{i\nu_1}\nu_1+V^N_{i\nu_2}\nu_2+V^N_{i\nu_3}\nu_3 \\[3pt] V^E_{ie}e+V^E_{i\mu}\mu+V^E_{i\tau}\tau\end{pmatrix},\qquad i=2,3
\label{mix}
\ee 
where the CKM and PMNS matrices are given by
$V^{CKM}=V^{U\dagger}V^D$ and $V_{PMNS}=V^{N\dagger}V^E$
respectively. Written in terms of the mass eigenstates, Eq.~(\ref{coup2}) generates the Lagrangian in Eq.~(\ref{coup1}). Note that the couplings
$\Gamma_{b,\mu}$ to the current eigenstates can be of $O(1)$ while
the coupling to mass eigenstates $\Gamma_i^m$ defined in the previous
section include the mixing angles and can be small. In particular the
relevant couplings in \chgdone{Eq.~(}{Eq.~(}\ref{coup1}) are given
by\footnote{ Here we assume the mixing is dominated by the down quark
  sector as is often the case in family symmetry models which relate
  the quark mass hierarchy to the quark mixing angles.} 
$\Gamma^m_s\chgdone{\approx}{=} \Gamma_b \chgdone{V_{cb}}{V^D_{3s}\approx\Gamma_b V^{CKM}_{cb}}$ and
$\Gamma^m_b=\Gamma_b V^D_{3b} \approx \Gamma_b$ and so the combination
$\Gamma_s^{m*}\Gamma^m_b\propto V^{CKM}_{cb}$ is small.

\subsection{Dark matter}
\label{subsec:DM}
It is of interest to ask if the new states could be the source of dark matter. For this to be the case it is necessary for the lightest state to be stable and this can readily be arranged through an additional $Z_2$ symmetry under which only the fields $\Psi_{Q,\ell}, \Phi$ are odd. Then these states can only be produced in pairs and the lightest of these states will be stable and a candidate for dark matter. The case there is no $Z_2$ symmetry is briefly discussed in App.~\ref{app:no-DM}.

\subsection{Fermion mass and mixing structure}
The $U(1)_F$ symmetry also controls the masses and mixing angles of the quarks and charged leptons when the symmetry is spontaneously broken. 
Quark and lepton masses are generated via the terms allowed by the $U(1)_F$ symmetry in the effective Lagrangian:
\bea
{\cal L}_{eff}^m&\approx&
\bar Q_{3,L} \;q_{3,R}\;H
+ \bar Q_{3,L} \;q_{2,R}\;H
+\bar Q_{2,L} \;q_{2,R}\; H\;({\chi\over M_q})^2
+ \bar Q_{2,L} \;q_{3,R}\; H\;({\chi\over M_q})^2\nonumber\\
&&+\bar L_{3,L}l_{3,R}H+\bar L_{2,L}l_{2,R}H{\chi\over M_q}+\bar L_{2,L}l_{3,R}H({\chi\over M_q})^4+\bar L_{3,L}l_{2,R}H({\chi^\dagger\over M_q})^3+h.c.
\eea
where we have suppressed the $O(1)$ coupling constants. Terms with $\Psi_{Q,L}$ replaced for $Q_2$ are forbidden by a $Z_2$
symmetry under which only the fields $\Psi_{Q,\ell}, \Phi$ are odd;
see Sec.~\ref{subsec:DM}. 
$M_q$ are
mediator masses\footnote{For simplicity we take the down quark and
  lepton mediators to have equal masses as might be justified by an
  underlying stage of Grand Unification. } associated with additional
vector-like states associated with the Froggatt Nielsen
mechanism.\footnote{We assume these are much heavier than the
  vector-like states, $\Psi_{Q,L}$ introduced in Eq.~(\ref{coup1}) so
  they do not contribute significantly B decays and mixing.} The
second generation quark masses are suppressed relative to the third generation masses by the factor $\epsilon^2=({<\chi>\over M_q})^2$ and to explain why this ratio is smaller in the up quark sector than in the down quark sector we need $M_u>M_d$. The same factor determines the mixing of the left-handed states giving $V^D_{2b}\sim{m_s\over m_b}$, $V^U_{2t}\sim {m_c\over m_t}$. Thus the dominant contribution to the $(2,3)$ element of the CKM matrix comes from the down quark sector giving  $V^{CKM}_{cb}\sim\epsilon^2\sim V^D_{2b}=O(\epsilon^2)\sim {m_s\over m_b}\sim 0.04$. 

The ratio of charged lepton masses is given by ${m_\mu\over m_\tau}=O(\epsilon)$.   The charged lepton mixing matrices are strongly constrained by the bounds on lepton flavour violation. For example, the branching ratio for $\tau\rightarrow \mu+\gamma$ is less than
$10^{-8}$ and an even stronger bound holds for $\mu\rightarrow
e+\gamma$. As discussed in Sec.~\ref{LFV}, to avoid violating these
bounds the charged lepton mixing angles must be very small.\footnote{This means that the large mixing angles observed in neutrino oscillation must come from the neutrino sector.} Here the choice of the $U(1)_F$ lepton charges 
ensures the mixing between the tau and muon is strongly suppressed and is of $O(\epsilon^4)$.  An additional $Z_3$ symmetry can eliminate charged lepton mixing completely as discussed in App.~\ref{app:noLepMix}.

What about the mass structure for the light generation?  A viable
structure  for the quark masses and mixing together
with a strong suppression of the mixing between the lepton doublets
can readily be obtained via the same $U(1)_F$ family symmetry through the
choice of charges given in Table \ref{MSAfirstF}, with the interactions
 \bea
{\cal L}_{eff}&=&\bar Q_1P_R\Psi_Q\Phi({\chi\over M_q})^3+\bar Q_{1,L}
\;q_{1,R}\;H({\chi\over M_q})^4+ \bar Q_{1,L} \;q_{2,R}\;H({\chi\over
  M_q})^3+\bar Q_{2,L} \;q_{1,R}\; H\;({\chi\over M_q})^3\nonumber\\
\label{eq:massLag}
&&+\bar L_{1,L}\Psi_{\ell,R}\Phi({\chi\dagger\over M_q})^6+\bar L_{1,L}l_{1,R}H({\chi\dagger\over M_q})^5+\bar L_{1,L}l_{2,R}H({\chi\dagger\over M_q})^5+\bar L_{2,L}l_{1,R}H({\chi\over M_q})+\text{h.c.}
\eea

\begin{table}
\[
\begin{array}{*{20}{c}}
{}&\vline& {{Q_1}}&{{q_1}}&{{L_1}}&{{l_1}}\\
\hline
Q_F&\vline& 3&-1&-4&1
\end{array}\]
\caption{\label{MSAfirstF}\it  \small  $U(1)_F$ family symmetry charges for the first family}
\end{table}

In the quark sector  the down quark mass is suppressed relative to the
strange quark masse by the factor $\epsilon^2$ and the Cabibbo angle
is of order $\epsilon\approx\sqrt{({m_d\over m_s})}$ giving a  realistic
mass and mixing angle structure. The additional coupling of $Q_1$ to
the vector-like quark will induce further lepton non-universal decays
at $O(\epsilon^3)$, the leading one being $b\rightarrow d\mu\mu$. As
is shown in App.~\ref{app:no-DM}  these can be eliminated by an extension of the family symmetry.

 In the charged lepton sector the electron mass is suppressed relative to the muon mass by the factor $\epsilon^4$ while the mixing between the muon and electron doublets is highly suppressed being also of $O(\epsilon^4)$.

\section{Flavour changing in the scalar sector}\label{newfc}
The model introduced in \chgdone{Eq.~(}{Eq.~(}\ref{coup2}) generates the flavour change in the quark sector via the quark mixing of \chgdone{Eq.~(}{Eq.~(}\ref{mix}). However it is also possible that the flavour change occurs in the scalar sector and to illustrate this we generalise the model slightly to include another Standard Model singlet scalar ${\tilde\Phi}$ with the interaction Lagrangian now given by  
\be
{\cal L}_{{\mathop{\rm int}} } =  {\Gamma_b\bar
  Q_3}{P_R}{\Psi_Q}{\Phi}  
+{\widetilde\Gamma_s\bar Q_2}{P_R}{\Psi_Q}{\tilde\Phi} + \Gamma_\mu\bar L_2 {P_R}{\Psi_\ell}\Phi+
{\rm{h}}{\rm{.c.}}
\label{coup3}
\ee
The inclusion of the scalar ${\tilde\Phi}$ has no implications for the
fermion mass and mixing model introduced in Sec.~\ref{family}, except for presence
of the additional coupling $\bar Q_1\Psi_Q(\tilde\Phi {\chi\over
  M_q})$ in Eq.~\eqref{eq:massLag}.

\subsection{Family symmetry properties of
  ${\tilde\Phi}$}\label{caveat}
The new coupling involving ${\tilde\Phi}$ is allowed by the $U(1)_F$ and $Z_2$ family symmetries provided  it is uncharged under $U(1)_F$ and odd under $Z_2$.    
 Indeed, it can be a real field or, if complex\footnote{In what follows
   we assume that $\tilde\Phi$ is complex.},  there is an additional
 coupling allowed in Eq.~(\ref{coup3}), namely  ${\widetilde\Gamma'_s\bar
   Q_2}{P_R}{\Psi_Q}{\tilde\Phi^*}$, corresponding to the second term
 with $\tilde\Phi$ replaced by $\tilde\Phi^*$. One can avoid this by
 extending the family symmetry to give $\tilde\Phi$ a family
 charge. If this is not the case  box diagrams involving an incoming
 and outgoing $\tilde\Phi$ coupled to the same fermion line will have
 additional crossed graphs with the $\tilde\Phi$ lines exchanged. For
 the case $\tilde\Phi$ is real the two graphs cancel exactly. For the
 case $\tilde\Phi$ is complex the combined contribution is
 proportional to $\widetilde\Gamma_s-\widetilde\Gamma_s'$.\footnote{We
   assume that, as in the models presented in Sec.~\ref{family} and
     in the Appendix,  $\Phi$ carries a family charge and
   so does not lead to further crossed graphs involving the $\Phi$
   vertex.}  
 The cancellation or partial cancellation of these graphs does not
 affect the order of magnitude estimate for the $b\to s\mu^+\mu^-$ and
 $B_s-\overline{B}_s$ mixing amplitudes presented in
 Sec.~\ref{fullamp} but does significantly change the expectation,
 discussed below, for $K-\bar K$ and $D-\bar D\;$mixing coming from the box diagram
 involving only $Q_2$ current quarks.

\subsection{Scalar mixing}

\begin{figure*}[t]
\begin{center}
\begin{tikzpicture}[scale=0.8]
\coordinate (S) at (7.3,0);
\coordinate (A) at (0,0);
\coordinate (B) at (0,1);
\coordinate (B2) at ($2*(B)$);
\coordinate (C) at (2,0);
\coordinate (D) at ($-2*(B)$);

\begin{scope}[very thick, every node/.style={sloped,allow upside down}]
\draw node[left] {$b_L$} (A) -- node {\midarrow}   ++(B2)  -- node {\midarrow}
++(B2)  -- node {\midarrow} ++ (B2) node[left] {$s_L$}; 
\node[left] at ($(A)+3*(B)$) {$\Psi_Q$};
\draw  ($(C)+6*(B)$) node[right] {$\mu_L$}-- node {\midarrow}   ++(D)  -- node {\midarrow}
++(D)  -- node {\midarrow}
++(D) node[right] {$\mu_L$} ; 
\node[right] at ($(C)+3*(B)$) {$\Psi_\ell$};
\draw[dashed] (B2) -- node {\midarrow}+(C);
\node[below] at ($0.5*(C)+(B2)$) {$\Phi_{L,H}$};
\draw[dashed] ($(C)+2*(B2)$) -- node {\midarrow} +($-1*(C)$);
\node[above] at ($0.5*(C)+2*(B2)$) {$\Phi_{L,H}$};

\node[below] at ($(A)+0.5*(C)$) {(a)};

\coordinate (A) at (S);
\coordinate (B) at (0,1);
\coordinate (B2) at ($2*(B)$);
\coordinate (C) at (2,0);
\coordinate (D) at ($-2*(B)$);
\coordinate (E) at ($(A)+(C)$);

\draw (A)  node[left] {$b_L$} -- node {\midarrow}   ++(B2)  -- node {\midarrow}
++(B2)  -- node {\midarrow} ++ (B2) node[left] {$s_L$}; 
\node[left] at ($(A)+3*(B)$) {$\Psi_Q$};

\draw  ($(E)+6*(B)$) node[right] {$b_L$} -- node {\midarrow}   ++(D)  -- node {\midarrow}
++(D)  -- node {\midarrow} ++(D) node[right] {$s_L$} ; 
\node[right] at ($(E)+3*(B)$) {$\Psi_Q$};

\draw[dashed] ($(A)+(B2)$) -- node {\midarrow} +(C);
\node[below] at ($(A)+0.5*(C)+(B2)$) {$\Phi_{L,H}$};
\draw[dashed] ($(A)+2*(B2)+(C)$) -- node {\midarrow} +($-1*(C)$);
\node[above] at ($(A)+0.5*(C)+2*(B2)$) {$\Phi_{L,H}$};
\node[below] at ($(A)+0.5*(C)$) {(b)};

\end{scope}
\end{tikzpicture}
\end{center}
\vspace{-1cm}
\caption{\it  \small Box-diagrams contributing to (a) $b\to s
  \mu^+\mu^-$ and  (b) $B_s\to \bar B_s$, with couplings given in Eq.~(\ref{coup3}).
\label{Box2} }
\end{figure*}
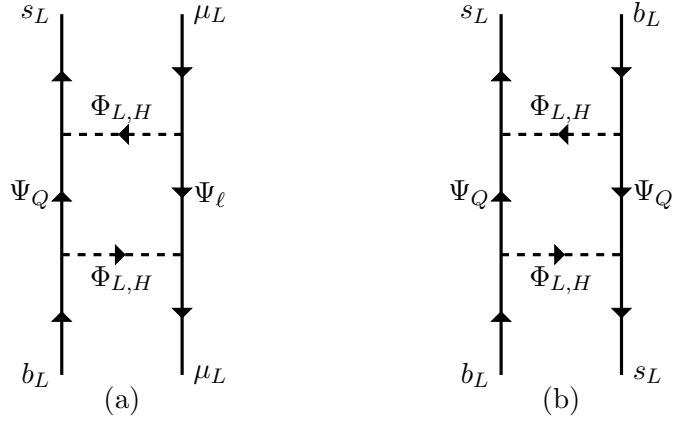

Since $\Phi\tilde\Phi\chi^2$ is an allowed scalar coupling  we expect $\Phi$ and $\tilde\Phi$ to mix so that the light and heavy $\Phi_L$ and $\Phi_H$ scalar mass
eigenstates  are mixtures of the ``current" scalar
states $\Phi$ and $\tilde\Phi$
\bea
\Phi_L&=&\cos(\theta)\;\Phi+\sin(\theta)\;\tilde\Phi\nonumber\\
\Phi_H&=&-\sin(\theta)\;\Phi+\cos(\theta)\;\tilde\Phi
\label{mixing}
\eea
In this case the leading contribution
to $b\to s\mu^+\mu^-$ is given by  Fig.~\ref{Box2}. Because $\Phi$ and
$\tilde\Phi$ couple to different quark current eigenstates there will
be a contribution to $B-\bar B$ mixing and $b\rightarrow s\mu\mu$ even
if the quark mixing angles are zero provided the scalar mass
eigenstates are mixture of the current eigenstates. \chgdone{}{In the
  limit that $\Phi$ and $\tilde\Phi$ do not mix, the Lagrangian
  \eqref{coup3} exhibits two separate U(1) symmetries $\Phi\to
  e^{i\alpha}\Phi$, $Q_3\to e^{i\alpha}Q_3$, $L_2\to e^{i\alpha}L_2$, $\tilde\Phi\to
  e^{i\beta}\tilde\Phi$, and $Q_2\to e^{i\beta}Q_2$, and the flavor
  mixing is all due to the matrices $V^{D,U,N,E}$ in Eq.~\eqref{mix}.} If these states are
  close in mass there will be a significant cancellation of the light
  and heavy  contributions.

\subsection{Scalar couplings in the mass eigenstate basis}\label{fullamp}
To simplify the presentation, here we concentrate on the case of
maximal mixing between $\Phi$ and $\tilde\Phi$,\footnote{Such maximal
  mixing is natural if the coefficient  of the
  quadratic term mixing $\Phi$ and $\tilde\Phi$ is much greater than the
  difference of the coefficients of the quadratic terms in $\Phi$
  and $\tilde\Phi$ separately.}, $\theta=\pi/4$, where now
$\Phi_{L,H}=(\Phi\pm\tilde\Phi)/\sqrt{2}$.  In Table{\ref{comparison1}) we give an example 
how the phenomenology changes when the states depart {from} maximal
mixing.

For the case that $\Phi$ and $\tilde\Phi$ are complex fields with only
the couplings of Eq.~(\ref{coup3}), the coupling of the maximally
mixed fields to mass eigenstates is given by 
\bea
\mathcal{L} &=& \frac{1}{{\sqrt 2 }}\left[
 {\left( {{\Gamma _b}V_{3b}^* +\chgdone{ \widetilde \Gamma _s^*}{ \widetilde \Gamma _s}V_{2b}^*}
   \right){\Phi _L} 
   + \left( {{\Gamma _b}V_{\chgdone{2}{3}b}^* - \chgdone{ \widetilde \Gamma _s^*}{ \widetilde \Gamma _s}V_{2b}^*}
   \right){\Phi _H}} \right]{\overline b _L}{\Psi _Q}\nonumber\\
&& + \left[ {\left( {{\Gamma _b}V_{3s}^* + \chgdone{ \widetilde \Gamma _s^*}{ \widetilde \Gamma _s}V_{2s}^*} \right){\Phi _L} + \left( {{\Gamma _b}\chgdone{V_{2s}}{V^*_{3s}} -
       \chgdone{ \widetilde \Gamma _s^*}{ \widetilde \Gamma _s}V_{2s}^*} \right){\Phi _H}}
\right]{\overline s _L}{\Psi _Q} + \text{h.c.}  
\eea 
Here and below $V_{ij}\equiv V_{ij}^D$. With these couplings it is
straightforward to determine the $B_s-\bar B_s$ mixing and
$b\rightarrow s\mu\mu$ amplitudes for the maximally mixed two scalar
model.

\subsection{$b\to s\mu^+\mu^-$}
The graph of Fig.~\ref{Box2}(a) gives the Wilson coefficients\\
 \bea\label{mumu1} 
C_9& =& - C_{10}\nonumber\\&=& -\frac{1}{{128\pi }{\alpha_{EM}}m_{\Psi_Q} ^2}
{\left| {{\Gamma _\mu }V_{2\mu }^E} \right|^2}
\bigg[\big( {\Gamma_b^*V_{3b}^{\phantom{*}} + \widetilde \Gamma _s^*V_{2b}^{\phantom{*}}} \big){\big(
    {{\Gamma _b ^{\phantom{*}}}V_{3s}^* + \widetilde \Gamma _s ^{\phantom{*}} V_{2s}^*}
  \big)}G\left( {y_\ell,{y_{{\phi _L}}},{y_{{\phi _L}}}} \right)
\nonumber\\&&+ 2\big( {{{\left| {{\Gamma _b ^{\phantom{*}}}} \right|}^2}V_{3b}^{\phantom{*}} V_{3s}^* -
    {\big| {\widetilde \Gamma }_s ^{\phantom{*}} \big|^2}V_{2b}^{\phantom{*}} V_{2s}^*}
\big)G\left( {y_\ell,{y_{{\phi _L}}},{y_{{\phi _H}}}} \right)
\nonumber\\&&+ \big( {\Gamma _b^*V_{3b}^{\phantom{*}} -  {\widetilde \Gamma }_s^*V_{2b}^{\phantom{*}}} \big)\big( {{\Gamma _b ^{\phantom{*}}}V_{3s}^* -  {\widetilde \Gamma }_s ^{\phantom{*}} V_{2s}^*} \big)G\left( {y_\ell,{y_{{\phi _H}}},{y_{{\phi _H}}}} \right)\bigg]
\nonumber\\
&\approx &-\frac{1}{128\pi \alpha_{EM}m_{\Psi_Q} ^2}
\left| \Gamma _\mu V_{2\mu }^E \right|^2
\bigg[\Gamma _b^*\chgdone{\Gamma _s}{\widetilde\Gamma _s}\left( G\left( y_\ell,y_{\phi _L},y_{\phi _L} \right) - G\left( y_\ell,y_{\phi _H},y_{\phi _H} \right) \right) \nonumber\\
&& \chgdone{+V_{3s}}{-V_{2b}}
\left( \big| \Gamma _b \big|^2 - \big|\chgdone{\Gamma _s}{\widetilde\Gamma _s}\big|^2 \right)\left( G\left( y_\ell,y_{\phi _L},y_{\phi _L} \right) +G\left( y_\ell,y_{\phi _H},y_{\phi _H} \right) \right)
\nonumber\\
&&\chgdone{+ 2V_{3s}}{-2V_{2b}}\left( \big| \Gamma _b \big|^2 
+ \big| \chgdone{\Gamma _s}{\widetilde\Gamma _s} \big|^2 \right)G\left( y_\ell,y_{\phi _L},y_{\phi _H}  \right)\bigg]
\label{C9}
\eea
\chgdone{}{In the last step we have used $\sum_iV_{ib}V^*_{is}=0$ and
  $V_{ij}\approx V^{CKM}_{ij}$ so that $V_{2s}\approx V_{3b}\approx1$
  and $V_{3s}\approx -V_{2b}^*$, with $|V_{2b}|\sim10^{-2}$.}

 The dimensionless function,  $G$, is a function of three mass
   ratios $G\equiv G(y_\ell,y_{\Phi_a},y_{\Phi_b})$ with
   $y_\ell=m_{\Psi_\ell}^2/m_{\Psi_Q}^2$,
   $y_{\Phi_a}=m_{\Phi_a}^2/m_{\Psi_Q}^2$,
   $y_{\Phi_b}=m_{\Phi_b}^2/m_{\Psi_Q}^2$ given
   by\footnote{Note that $F(x,x)\equiv x^{-1} G(1,x^{-1},x^{-1})$.}\bea
G(y_\ell,y_{\Phi_a},y_{\Phi_b})&=&{y_\ell^2 \ln(y_\ell)\over (y_\ell-1)(y_\ell-y_{\Phi_a})(y_\ell-y_{\Phi_b})}+{y_{\Phi_a}^2 \ln(y_{\Phi_a})\over (y_{\Phi_a}-1)(y_{\Phi_a}-y_\ell)(y_{\Phi_a}-y_{\Phi_b})}\nonumber\\
&&+{y_{\Phi_b}^2 \ln(y_{\Phi_b})\over (y_{\Phi_b}-1)(y_{\Phi_b}-y_\ell)(y_{\Phi_b}-y_{\Phi_a})}
\label{bsmm}
\eea

We see that there are two distinct contributions to $b\to
s\mu^+\mu^-$. The term third line from the bottom of \chgdone{Eq.~(}{Eq.~(}\ref{mumu1}) comes from the scalar mixing while the remaining terms, proportional to the small mixing angle $V_{3b}$, come from quark mixing. To
 avoid the mixing angle suppression the contribution from scalar mixing should
 be large which in turn requires that the mass of $\Psi_\ell$ should not be significantly greater than mass difference, $\delta m$, between the masses of $\Phi_H$ and $\Phi_L$. 

 \subsubsection{$B_s-\overline{B}_s$ mixing}
 \label{ssec:BB}
The graph of Fig.~\ref{Box2}(b)  gives
 \begin{multline}
C_{B\bar B}={1\over 512\pi^2m_{\Psi_Q}^2}
\left[ {\left( {\Gamma _b^*V_{3b}^{\phantom{*}} + \widetilde \Gamma _s^*V_{2b}^{\phantom{*}}} \right)^2}{\left( {{\Gamma _b ^{\phantom{*}}}V_{3s}^* + \widetilde \Gamma _s ^{\phantom{*}} V_{2s}^*} \right)^2}G\left( {1,{y_{{\phi _L}}},{y_{{\phi _L}}}} \right)\right.\\
 + 2\left( {{{\left(\vphantom{\widetilde \Gamma _b^*} {\Gamma _b^*V_{3b}^{\phantom{*}}} \right)}^2} - {{\left(
           {\widetilde \Gamma _s^*V_{2b}^{\phantom{*}}} \right)}^2}} \right)
\left( {{{\left(\vphantom{\widetilde \Gamma _b^*}  {{\Gamma _b^{\phantom{*}}}V_{3s}^*} \right)}^2} - {{\left( {\widetilde \Gamma _s^{\phantom{*}}V_{2s}^*} \right)}^2}} \right)G\left( {1,{y_{{\phi _L}}},{y_{{\phi _{H}}}}} \right)\\
\left. + {\left( {\Gamma _b^*V_{3b}^{} - \widetilde \Gamma _s^*V_{2b}^{}} \right)^{{2^{}}}}{\left( {{\Gamma _b^{\phantom{*}}}V_{3s}^* - \widetilde \Gamma _s^{\phantom{*}}V_{2s}^*} \right)^2}G\left( {1,{y_{{\phi _H}}},{y_{{\phi _H}}}} \right)\right]
\label{BBbar1}
\end{multline}
In the same limit discussed above this is approximately given by
\begin{multline}
C_{B\bar B} \approx {1\over 512\pi^2m_{\Psi_Q}^2}\bigg[
{{\big( {\Gamma _b^*{{\widetilde \Gamma }_s}} \big)^2}\big( {G\left( {1,{y_{{\phi _L}}},{y_{{\phi _L}}}} \right) + G\left( {1,{y_{{\phi _H}}},{y_{{\phi _H}}}} \right) - 2G\left( {1,{y_{{\phi_L}}},{y_{{\phi _H}}}} \right)} \big)} 
\\
 - 2{V_{2b}}\Gamma _b^*{\widetilde \Gamma _s}\Big( {{{\big| {{\Gamma _b}} \big|}^2} - {{\big| {\widetilde\Gamma}_s \Big|}^2}} \big)\big( {G\left( {1,{y_{{\phi _L}}},{y_{{\phi _L}}}} \right) - G\left( {1,{y_{{\phi _H}}},{y_{{\phi _H}}}} \right)} \big)
\\
+ (V_{2b})^2\Big [
2\Big( {{{\big| {{\Gamma _b}} \big|}^2} - {{\big| {{\widetilde\Gamma}_s} \big|}^2}}
  \Big)^2\big( G( {1,{y_{{\phi _L}}},{y_{{\phi _L}}}} ) + G( {1,{y_{{\phi
                _H}}},{y_{{\phi _H}}}} ) \big) 
\\
{-\Big( \big| {{\Gamma _b}} \big|^4 + \big| {{\widetilde\Gamma}_s} \big|^4
  \Big) \big( {G\left( {1,{y_{{\phi _L}}},{y_{{\phi _L}}}} \right) + G\left( {1,{y_{{\phi _H}}},{y_{{\phi _H}}}} \right) - 2G\left( {1,{y_{{\phi_L}}},{y_{{\phi _H}}}} \right)} \big)}
\Big ]+\cdots\bigg]\,,
\label{bbnew}
\end{multline}
where the ellipsis stand for terms of order $(V_{2b})^3$.

For the case $\delta m\chgdone{}{=m_{\Phi_H}-m_{\Phi_L}}\ll m_{\Psi_Q}$ the first term, unsuppressed by
powers of $V_{2b}$, is suppressed by \chgdone{$\delta m^2\over m_{\Psi_Q}^2$}{$\delta m^2/ m_{\Psi_Q}^2$}. \chgdone{}{The second term,
  suppressed by a single power of $V_{2b}$, is also suppressed by \chgdone{}{$\delta m^2/m_{\Psi_Q}^2$} and by 
  $\delta\Gamma/\Gamma=(|\Gamma_b|-|\Gamma_s|)/(|\Gamma_b|+|\Gamma_s|)$. Finally the terms on
the third and fourth line of Eq.~\eqref{bbnew}, both suppressed by $V_{2b}^2$, are additionally
suppressed by  $(\delta\Gamma/\Gamma)^2$ and $\delta m^2/m_{\Psi_Q}^2$, respectively.} This demonstrates
how in this limit the bound on $\Gamma_{B_s-\overline{B}_s}$ coming from $B_s-\overline{B}_s$ mixing
does not significantly constrain $\Gamma_{b\to s\mu^+\mu^-}$\chgdone{ if the term in
Eq.~\chgdone{(\ref{mumu2})}{(\ref{mumu1})} is the dominant one it does not have this suppression if
$m_{\Psi_\ell}$ is not significantly greater than $\delta m$.}{; for example,  in the combined limit
$\delta m\to 0$ and $\delta\Gamma\to0$, $C_{B\overline{B}}$ vanishes \chgdone{is of order
$V_{2b}^3$} while $C^{\rm box}_{9,10}$ remain of order $V_{2b}$ .}

 To summarise, for the case $m_{\Psi_Q}\gg m_L\sim m_{\Phi_L}\sim
 m_{\Phi_H}$, there are two main differences between the contributions coming from quark mixing and coming from scalar mixing.  In the former the $b\to
 s\mu^+\mu^-$ amplitude is suppressed by a factor of $V_{2b}$ while in
 the latter there is no such overall suppression. Secondly, the relative
 suppression of the $B_s-\overline{B}_s$ mixing amplitude to the  $b\to s\mu^+\mu^-$ amplitude is $V_{2b}$ in the former and  \chgdone{$\delta m^2\over m_{\Psi_Q}^2$}{$\delta m/ m_{\Psi_Q}$}  in the latter.  Indeed combining the two contributions it is possible to eliminate the  $B_s-\overline{B}_s$ mixing bound entirely.

\section{Phenomenological analysis}\label{pheno}
In this Section we perform a numerical evaluation of the decay rates for the relevant processes and compare to the present measurements. For the case the box diagrams involve only a single scalar field, $\Phi$, all the flavour changing comes from quark mixing and we assume that this is dominated by the down quark sector. We will refer to this as the ``Single scalar model".

For the case that the box diagrams involve both the $\Phi$ and
$\tilde\Phi$ fields --- the ``Two scalar model" --- the flavour changing
effects come from both the down quark mixing and from the scalar
mixing. To illustrate the expectation in this model we will usually
present results for a benchmark point   with  the light scalar, $\Phi_L$, and the vector-like lepton having mass of
$100\;\text{GeV}$ and the heavy scalar, $\Phi_H$, with mass of $400\;\text{GeV}$ assuming down quark mixing is dominant. To emphasise that the scalar mixing can be significant we also present results assuming that the down quark
mixing is negligible together with the result assuming that the CKM mixing comes entirely from the down quark sector. Finally, to facilitate comparison between models, we quote the result for the benchmark choice,  $\Gamma_b=1.5$.

For most observables the results are quite insensitive to the precise
value of the light scalar mass. However, when discussing the  dark
matter and muon anomalous magnetic moment bounds, we will choose
$m_L=105\;\text{GeV}$ and $m_{\Phi_L}=62\;\text{GeV}$ to compare to the detailed
analysis of  Refs.~\cite{Kowalska:2017iqv , Calibbi:2018rzv}.
\subsection{$K_L-K_S$ mass difference and $D-\bar{D}$ mixing.}
\subsubsection{Single scalar model}
For the  $K_L-K_S$ mass difference and $D-\bar{D}$ mixing the Hamiltonian has the form
\bea
{\cal H}_{\rm eff}^{K_LK_S}=
 C_{K_LK_S}
 ({\bar s}_{\alpha} \gamma^{\mu} P_{L} d_{\alpha})\, ({\bar s}_{\beta} \gamma^{\mu} P_{L} d_{\beta})     \,,
 \\
  {\cal H}_{\rm eff}^{D\bar D}=C_{D\bar{D}}
 ({\bar c}_{\alpha} \gamma^{\mu} P_{L} u_{\alpha})\, ({\bar c}_{\beta} \gamma^{\mu} P_{L} u_{\beta})  \eea
where the $2\sigma$ bound is given by
\bea
C_{K_LK_S}=9\times 10^{-7}\;\text{TeV}^{-2}\\
C_{DD}=2.7\times 10^{-7}\;\text{TeV}^{-2}
\eea
 The calculation of the box diagrams are the same as for $B_s-\bar
 B_s$ mixing with the dominant term coming from the quark mixing of
 the box involving only $Q_3$ current quarks. As a result the
 coefficients $C_{K_LK_S} (C_{D\bar D})=\frac{\Gamma_{K_LK_S}\;
   (\Gamma_{D\bar D})} {128\pi^2m_{\Psi_Q}^2}$  for the single scalar
 model are given by
  \bea
\label{KKbound}
\Gamma_{K_LK_S} \approx (V^{D*}_{3d}V^D_{3s})^2|\Gamma_b|^4\\
\label{DDbound}
\Gamma_{D\bar D}\approx (V^{U*}_{3u}V^U_{3c})^2|\Gamma_b|^4
\eea 
Due to the small mixing angles this is so heavily suppressed that the
$K_L-K_S$ mass difference and $D-\bar{D}$ mixing  do not provide a bound on the coupling $\Gamma_b$. 
\subsubsection{Two scalar model}
The situation is quite different for the two scalar model. If the down quark mixing dominates 
the $K_L-K_S$ mass difference gives the most stringent bound on the coupling $\widetilde\Gamma_s$
of $\tilde\Phi$ to the second current quark family because the mixing angle to the first family is large with $V^D_{2d}\approx V^{CKM}_{cd}$.  Thus in the two scalar model lepton universality violation effects are  expected to be larger than in the single scalar case.

In the benchmark model,  $m_{\Psi_\ell} = m_{\Phi_L}
= 100\;\text{GeV}$, 
$m_{\Phi_H} = 400\;\text{GeV}$ and $m_{\Psi_Q} = 2.0\;\text{TeV}$, with $V^D_{2s}\approx1.0$, $V^D_{3d}\approx0$,
$V^D_{2d}\approx0.23$ and $V^D_{3s}\approx0.042$, the $K_L - K_S$ mass
difference is dominated by the $|\widetilde\Gamma_s|^4$ term  and 
we find the bound  $|\widetilde\Gamma_s|<0.57$.\footnote{Note that suppressing the $V^D_{2d}$ contribution  to the Cabibbo angle is not viable  because of the complementarity of the bounds
  from $K-\bar{K}$  and  $D-\bar{D}$ mixing.}
For the  $D - \bar D$ mixing case the bound is slightly stronger but can
readily be satisfied with the same $\widetilde\Gamma_s$ if the up sector
mixing angle is smaller, $V^U_{2u}<0.5\;V^{CKM}_{cd}$. Note that these
bounds may be evaded if there is a significant cancellation between
the contributing box graphs ({\it c.f.} Sec.~\ref{caveat}).

\subsection{\chgdone{$B-\bar B$}{$B_s-\bar B_s$} mixing}

For ease of comparison we will use the $2\sigma$ bounds quoted in \cite{Arnan:2016cpy} 
\be
C_{B\bar B}(\mu_H) \in [-2.1,0.6]\times 10^{-5}\, \text{TeV}^{-2}\quad (\text{at } 2\, \sigma),\\
\ee

\begin{figure}[t]
\begin{center}
\includegraphics[width=3.3in]{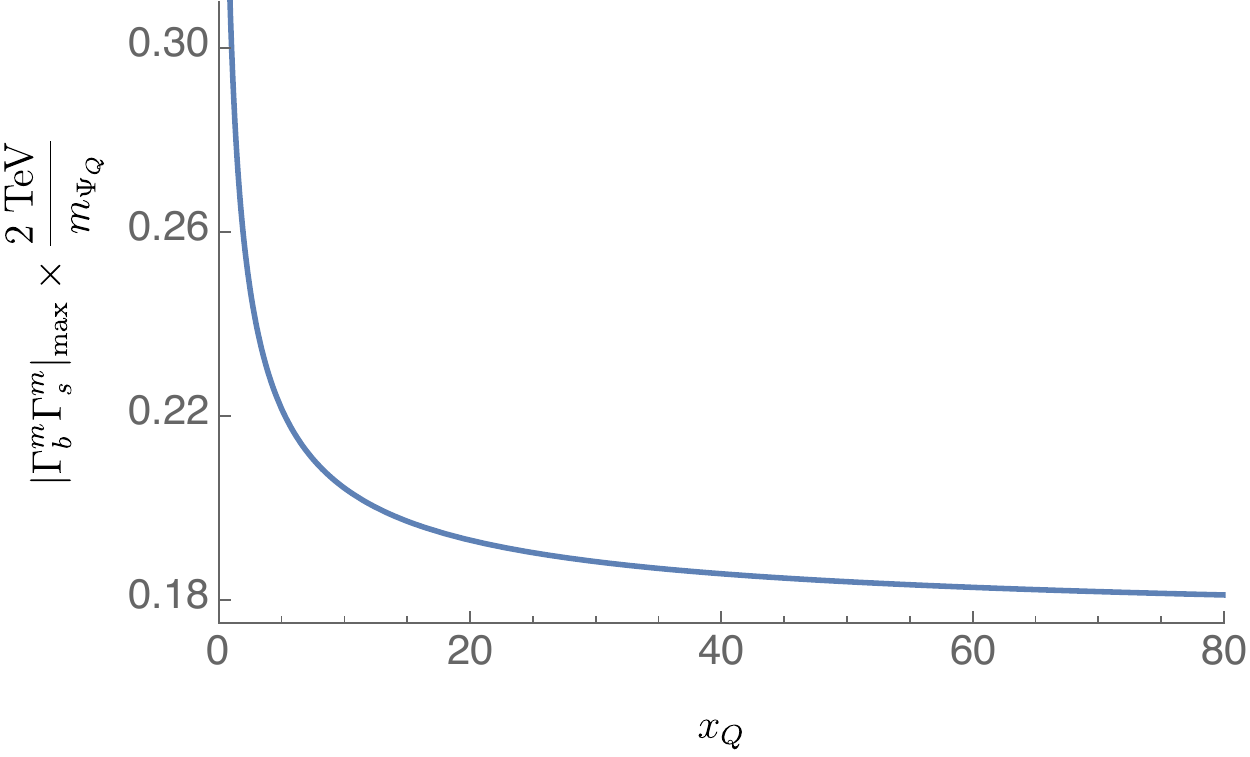}
\includegraphics[width=3.3in]{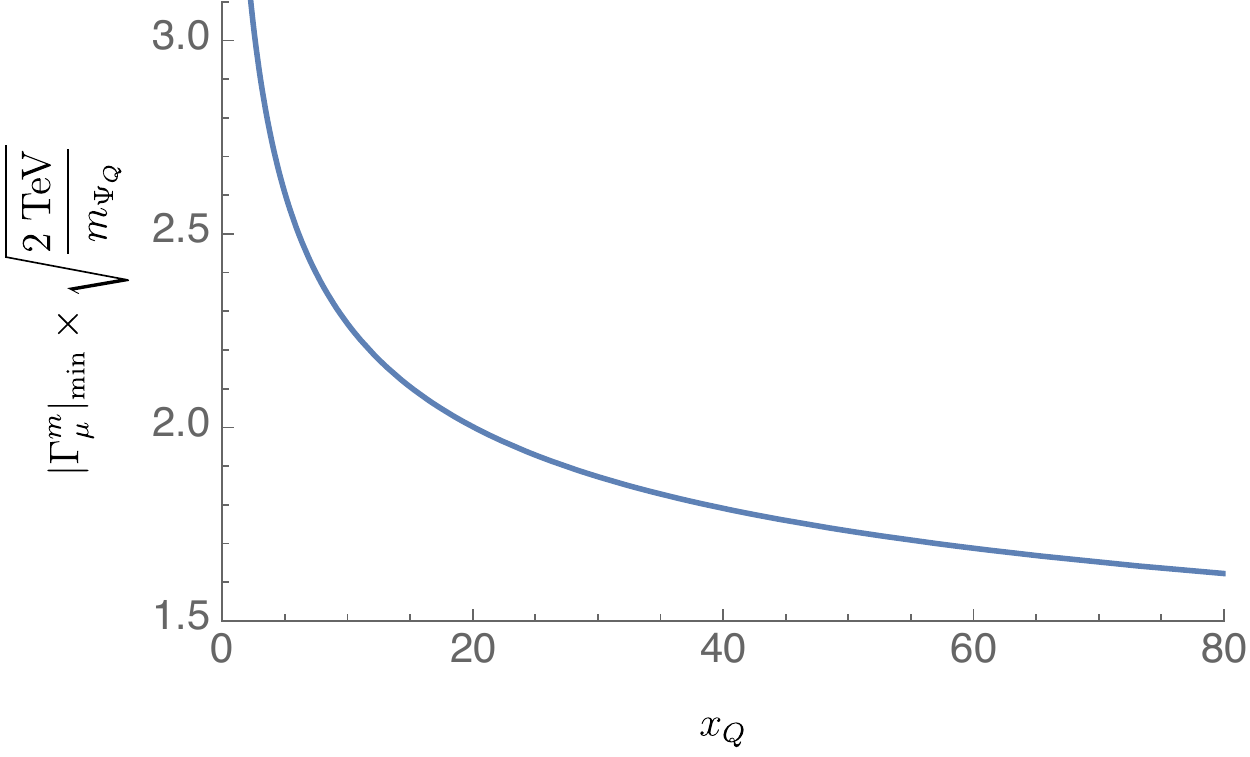}
\end{center}
\caption{\label{fig:singleSbounds} \it  \small  Comparison of the
  upper bound, for the simple single scalar model introduced in
  Sec.~\ref{intro},  on $|\Gamma_b^m\Gamma_s^m|$  from $B_s-\bar B_s$ mixing (left panel) with the lower bound on $|\Gamma_\mu^m|$ from $b\to s \mu\mu$ for $x_\ell=1$ (right panel), as a function of  $x_Q$.}
\end{figure}

\subsubsection{Single scalar model}

For the simple single scalar model introduced in Sec.~\ref{intro} the bound for  the degenerate mass case, $x_Q=x_\ell=1$, is given by
\begin{equation}
{|\Gamma^m_b\Gamma^m_s| \approx |V_{3b}^{\phantom{*}}V^*_{3s}||\Gamma_b|^2\le \:0.30\:\;\frac{m_\Psi}{2\,{\text{TeV}}}\,.}
\label{bd1}
\end{equation}
which gives
\begin{equation}
|\Gamma_b |\le \:2.6\:\;\sqrt{\frac{m_\Psi}{2\,{\text{TeV}}}}
\end{equation}

To illustrate the sensitivity to the  mass spectrum, we will consider
the extreme case, $x_Q\approx 400,\;x_\ell\approx1$ in which the scalar
and vector-like lepton are as light as possible, of order 100\;GeV,
consistent with present bounds, while the vector-like quark is of
order 2\;TeV to be consistent with LHC bounds. In this case the bound
is strengthened slightly giving 
\begin{equation}
{|\Gamma^m_b\Gamma^m_s |\le \:0.18\:\;\frac{m_\Psi}{2\,{\text{TeV}}}\,.}
\label{bs2}
\end{equation}
which gives
\begin{equation}
|\Gamma_b |\le \:2.0\:\;\sqrt{\frac{m_\Psi}{2\,{\text{TeV}}}}
\label{bs2again}
\end{equation}

The left panel in Fig.~\ref{fig:singleSbounds} shows this upper bound
as a function of $x_Q$. The bound is given at
$m_{\Psi_Q}=2\;\text{TeV}$, but other values are obtained by the
indicated scaling, $m_{\Psi_Q}/2\;\text{TeV}$.
 
\subsubsection{Two scalar model}

\begin{figure}[t]
\begin{center}
\includegraphics[width=3in]{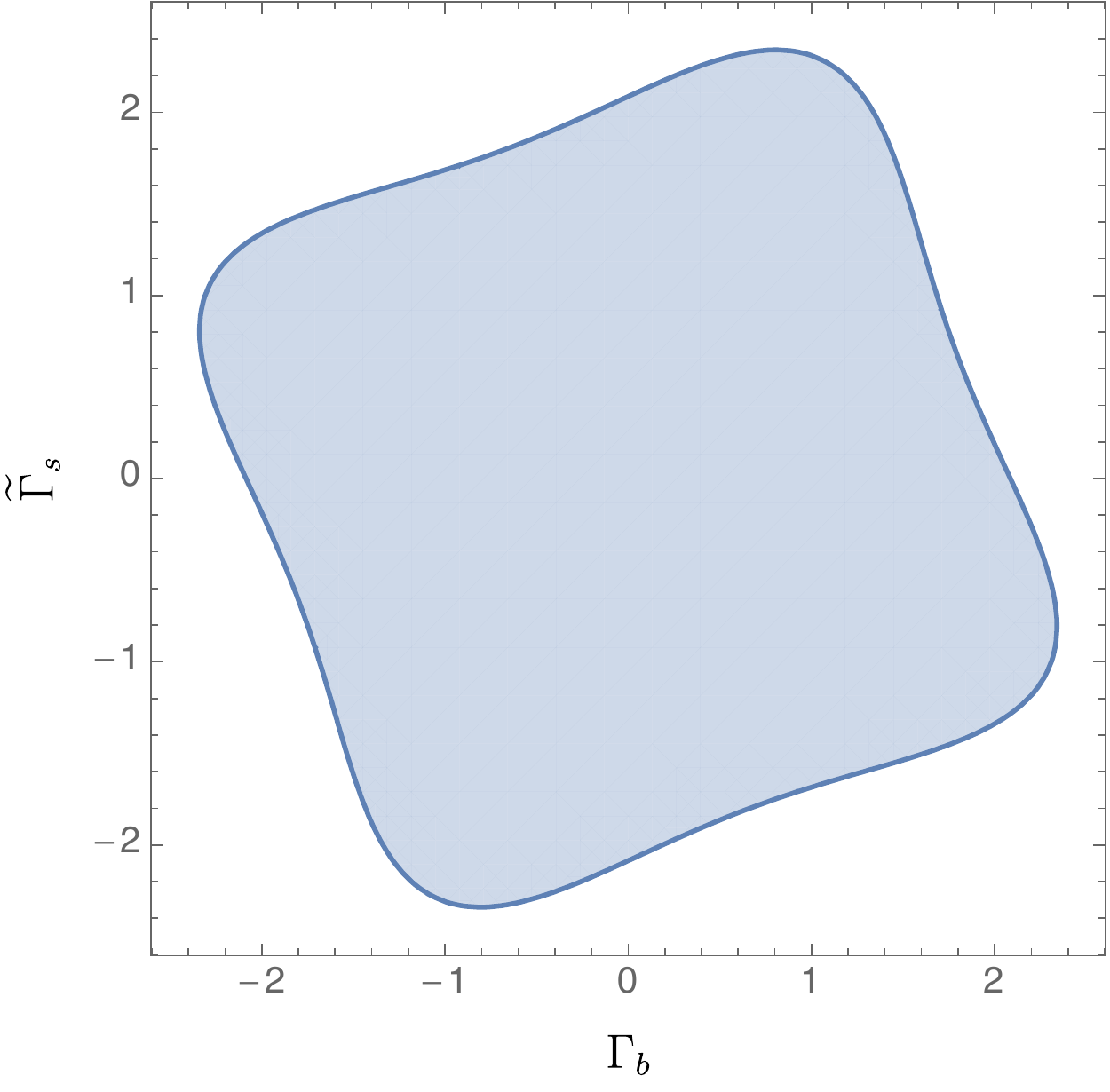}~~~
\includegraphics[width=3in]{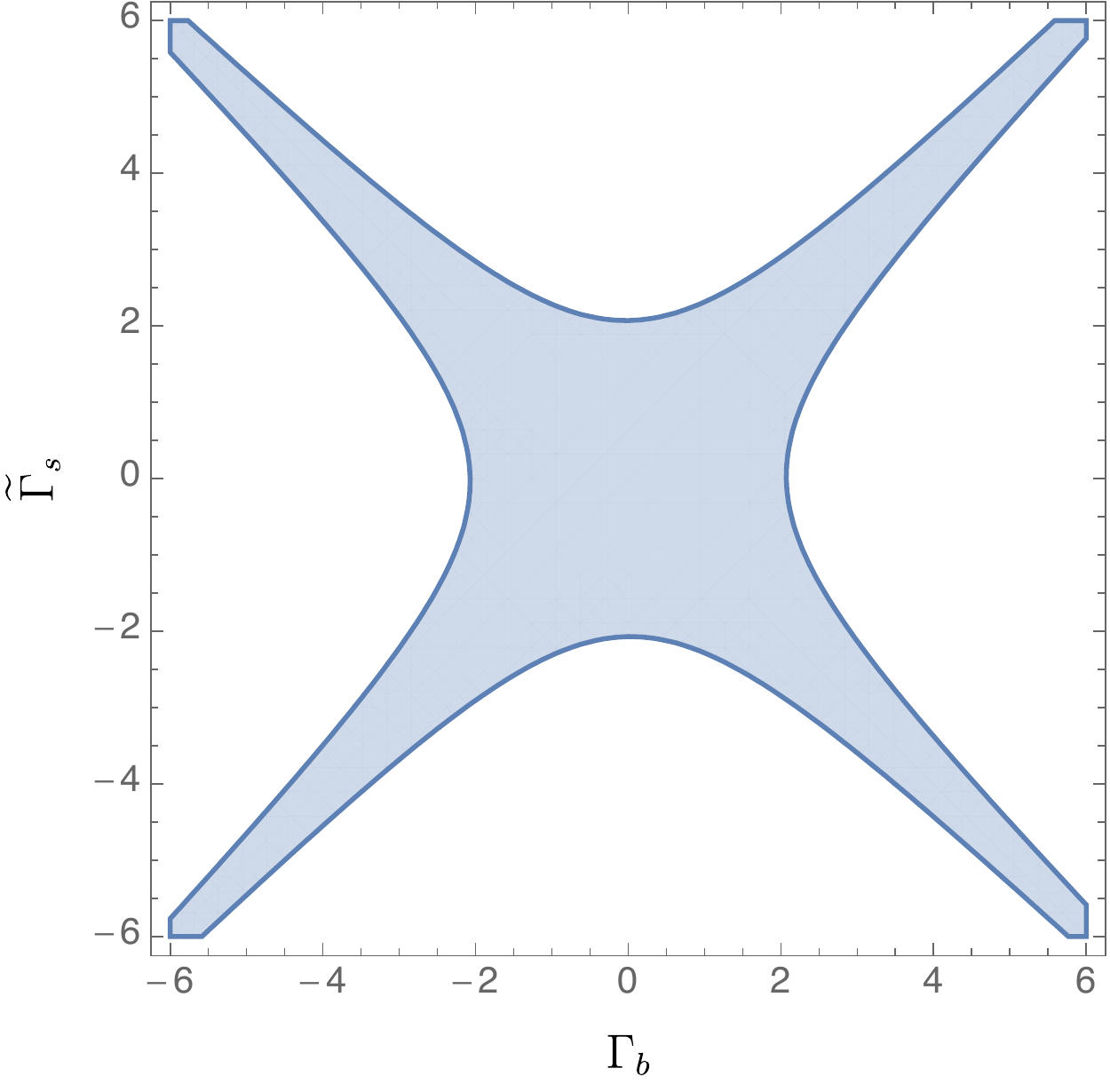}
\end{center}
\caption{\label{fig:BB2sm} \it  \small Allowed region (shaded)  in the 
  $\widetilde\Gamma_s$ vs $\Gamma_b$ plane at $2\sigma$ in the two scalar model from $B_s-\bar
  B_s$ mixing.  Left panel: using Eq.~\eqref{bbnew},  taking $m_{\Phi_L}=100\;\mathrm{GeV}$,
  $m_{\Phi_H}=400\;\mathrm{GeV}$, and $m_{\Psi_Q}=2\;\mathrm{TeV}$, and
  $V^{D*}_{3s}=-V^D_{2b}=0.042$. Right panel:  the equal mass limit at
  $m_{\Phi_L}=m_{\Phi_H}=200\;\mathrm{GeV}$, with $m_{\Psi_Q}=2\;\mathrm{TeV}$ and
  $V^{D*}_{3s}=-V^D_{2b}=0.042$;  the
  absence of a bound in the simultaneous $\delta m\to0$ and
  $\delta\Gamma\to0$ limits is evident.}
\end{figure}

 As discussed above the suppression of the $B-\bar B$ amplitude requires that
the scalar mass difference, $\delta m$ be small relative to $\chgdone{m_Q}{m_{\Psi_Q}}$ as in our benchmark model. The bound on $\Gamma_b$ depends on the mixing angle $V^D_{3s}$ and, for the extreme choices of this angle our benchmark model gives: 

\noindent
{\bf Down quark mixing vanishes, $V^D_{3s}=0$.}
Using the benchmark point $m_\ell = m_L
= 100\;\text{GeV}$, 
$m_H = 400\;\text{GeV}$ and $m_\Psi = 2\;\text{TeV}$ we find
\be
\label{eq:CBBnoV}
C_{B\bar B} \approx 0.022\frac{(\Gamma _b^*\widetilde \Gamma_s)^2}{{512{\pi ^2}m_{\Psi_Q}^2}}\,,
\ee
Taking  $|\widetilde\Gamma_{s}|=0.57$, { as needed for suppressing the
  $K_L-K_S$ mass difference, this gives, at the $2\sigma$ level,
 $
|\Gamma_b|<4.1
$.

\noindent
{\bf Down quark mixing maximal, $V^D_{3s}=V^{CKM}_{cb}$.}
The left panel of Fig.~\ref{fig:BB2sm} plots the $2\sigma$-allowed
region (shaded)  in the $\widetilde\Gamma_s$ {\it vs} $\Gamma_b$
plane, for $m_{\Phi_L}=100\;\text{GeV}$,
  $m_{\Phi_H}=400\;\text{GeV}$, and $m_{\Psi_Q}=2\;\text{TeV}$, and
  $V^{D*}_{3s}=-V^D_{2b}=0.042$. Taking  $|\widetilde\Gamma_{s}|=0.57$, we find the bound is satisfied at the $2\sigma$ level by 
$
|\Gamma_b|<1.8
$

Finally, to demonstrate the suppression discussed in
Sec.~\ref{ssec:BB} when $\delta\Gamma\to0$,  the right panel of Fig.~\ref{fig:BB2sm} shows that
  the bounds disappear in this limit. The region shown is for
  the equal mass limit at
  $m_{\Phi_L}=m_{\Phi_H}=200\;\text{GeV}$, with $m_{\Psi_Q}=2\;\text{TeV}$ and
  $V^{D*}_{3s}=-V^D_{2b}=0.042$. This limit is viable if there is a strong
  cancellation in $K^0 - \bar K^0$ and $D -\bar D$ mixing. 

\subsection{\Large $b\rightarrow s\mu\mu$}
\chgdone{Again we will}{We}  use  the $2\sigma$ bound quoted in \cite{Arnan:2016cpy} 
\be
-1.28\;\text{TeV}^{-2}\leq C_9=-C_{10}\leq	-0.49\;\text{TeV}^{-2}\,\quad (\text{at }2\,\sigma).
\label{2sigmamu}
\ee

\subsubsection{Single scalar model}
In the degenerate mass case the function $F(x_Q,x_\ell)=1/3$ in
\chgdone{Eq.~(}{Eq.~(}\ref{mumu}) and, for $m_\Phi=2\;\text{TeV}$, the
coupling is uncomfortably large, close to the nonperturbative limit,
$\left| \Gamma_\mu\right|\geq 3.8.
$
However the bound is significantly weakened in the extreme case
$x_Q\approx 400,\;x_\ell\approx1$  with
 $F(x_Q,x_\ell)=1.1\times 10^{-2}$ giving
$
1.3\leq\left| \Gamma_\mu\right|\leq 2.2
$,
for the largest possible value of $|\Gamma_b^m\Gamma_s^m|$ given in
Eq.~(\ref{bs2})  or equivalently for $|\Gamma_b|=2$ as given in Eq.~(\ref{bs2again}). To facilitate comparison between models we also quote
the result for the benchmark choice,  $\Gamma_b=1.5$, which gives 
the bound  
$
2.2>\left| \Gamma_\mu\right|\geq 1.8.
\label{bs5}
$

\subsubsection{Two scalar model}

\noindent
{\bf Down quark mixing vanishes, $V^D_{3s}=0$. } 
For the same benchmark point used in obtaining Eq.~\eqref{eq:CBBnoV},
this pure scalar mixing case gives
\be
C_9^{{\rm{box}}} = - C_{10}^{{\rm{box}}} \approx -\frac{2.1\;\Gamma _b^*\widetilde\Gamma_s \left |\Gamma _\mu \right|^2}{{128\pi }{\alpha_{EM}}m_{\Psi_Q} ^2} 
\label{mumunum}
\ee
For the benchmark choice $|\Gamma_b|=1.5$, $|\widetilde\Gamma_s|=0.57$ the bound on $\Gamma_{\mu}$ is given by
$
2.9>\left| \Gamma_\mu\right|\geq 1.8.
\label{bd5}
$

\noindent
{\bf Down quark mixing maximal, $V^D_{3s}=V^{CKM}_{cb}$.} 
For the benchmark point we find $
2.2>{\left|\Gamma_\mu\right|}\geq 1.4.
$
\subsection{Anomalous magnetic moment of the muon}
\label{subSec:amu}
The effective Hamiltonian has the form
\begin{equation}
{\cal H}_{\rm eff}^{a_\mu}={-}a_\mu \dfrac{e}{4 m_\mu} 
 (\bar{\mu}\sigma^{\lambda\nu} \mu)\, F_{\lambda \nu},
\end{equation}
where $a_\mu \equiv (g-2)_\mu/2$ is the anomalous magnetic moment  of the muon. 
\subsubsection{Single scalar model}
The one loop graph, Fig.~4 of \cite{Arnan:2016cpy}, gives a contribution to it of the form
\begin{equation}
\Delta a_\mu={m_\mu^2 |\Gamma_{\mu}|^2\over 8\pi^2}{ F_7(x_\ell)\over m_{\Phi}^2}\,,
\end{equation}
where
\begin{equation}
F_7(x)=\frac{ x^3-6 x^2+6 x \log x+3 x+2}{12 (x-1)^4}\,.
\end{equation}

In the limit of equal masses
\be
\Delta a_\mu=5.9\times 10^{-12} |\Gamma_{\mu}|^2\left(\frac{1\;\text{TeV}}{m_{\chgdone{\phi}{\Phi}}}\right)\,.
\ee
negligible when compared to the discrepancy of the Standard Model prediction from the experimental measurement
\begin{equation} \label{eq:g-2}
 \Delta a_\mu = a_\mu^\mathrm{exp}-a_\mu^\mathrm{SM} = (2.36\pm 0.87)\times 10^{-9}.
\end{equation}
Clearly $ \Delta a_\mu $ increases significantly in the non-degenerate
case with light $\Psi_\ell$ and $\Phi$. We will consider the
benchmark point  $m_{\Psi_\ell}=105\;\text{GeV}$ and
$m_{\Phi}=62\; \text{GeV}$.\footnote{This is taken
  because, even in the case {$\Phi$} is stable, this is an
  experimentally allowed point, {\it c.f.} Sec.~\ref{DM}.} To reduce the discrepancy of $a_{\mu}$ from 2.7$\sigma$ to 1$\sigma$ requires $\Gamma_\mu=1.4$ while to remove the discrepancy completely requires  $\Gamma_\mu=1.8$. It is noticeable that these values are not far from the bounds derived above.
\subsubsection{Two scalar model}
The anomalous magnetic moment does not involve flavour mixing so one gets a unique prediction.
In this case 
\begin{equation}
\Delta a_\mu={m_\mu^2 |\Gamma_{\mu}|^2\over 16\pi^2}\left({ F_7(x_{\Phi_L})\over m_{\Phi_L}^2}+{ F_7(x_{\Phi_H})\over m_{\Phi_H}^2}\right)\,,
\end{equation}
where $x_{\Phi_{L,H}}=m_{\Psi_\ell}^2/m_{\Phi_{L,H}}^2$.
 For the benchmark point $m_{\Psi_\ell}=105\;\text{GeV},\;m_{\Phi_L}=62\; \text{GeV},\; m_{\Phi_H}=400\;\text{GeV}$,
to reduce the discrepancy to 1$\sigma$ requires $\Gamma_\mu=1.9$ and to eliminate the discrepancy completely requires $\Gamma_\mu=2.4$.
The contribution is sensitive to the mixing angle, $\theta$, associated with the scalar mass eigenstates and reducing this from the maximal mixing value, $\theta=\pi/4$, to $\theta=\pi/8$ the value needed change to 1.5 and 1.9 respectively.

\begin{table}[t]
\begin{center}
\begin{tabular}{|c|c|c|c|}
\hline
  &\multirow{2}{*}{Single scalar}&\multicolumn{2}{|c|}{Two scalar }\\
  \cline{3-4}
&&$V^D_{2b}=0$&$V^D_{2b}=-0.043$\\
\hline
$ B_s - {\overline B}_s$&$ \Gamma _b < 2.0$& $\Gamma _b < 4.1\; (5.8)$& ${{\Gamma _b} < 2.3 \;(2.3)}$\\
\hline
  $ {b \to s\mu \mu \;}$& ${3>{\Gamma _\mu } > 1.8}$&
                                                      ${2.9\; (3.0)>{\Gamma _\mu } > 1.8\;(1.9)}$&${2.2\; (2.2)>{\Gamma _\mu } > 1.4\; (1.3)}$\\
\hline
$ (g - 2)_\mu $&$ {{\Gamma _\mu } = 1.4}$& ${{\Gamma _\mu } = 1.9\; (1.5)}$&${{\Gamma _\mu } = 1.9\; (1.5)}$ \\
\hline
 \end{tabular}
 \caption{\it \small Comparison of the parameters needed for different
   models.  In all cases we used $V^D_{2s}\approx1.0$,
   $V^D_{3d}\approx0$, $V^D_{2d}\approx0.23$.  For the single scalar
   model we have used $m_{\Psi_\ell} = m_{\Phi} =100\,\mathrm{GeV}$,
   and $m_{\Psi_Q} = 2.0\,\mathrm{TeV}$.  For both two-scalar models
   we used $m_{\Psi_\ell} = m_{\Phi_L} = 100\,\mathrm{GeV}$,
   $m_{\Phi_H} = 400\,\mathrm{GeV}$ and
   $m_{\Psi_Q} = 2.0\,\mathrm{TeV}$, and $|\widetilde\Gamma_s|=0.57$,
   that satisfies the bounds from $K_L-K_S$ mass difference. In the
   two-scalar models, $b\to s\mu\mu$ is computed at the benchmark
   point $|\Gamma _b| = 1.5$. In the last row the parameter values
   given reduce the discrepancy in $(g-2)_\mu$ from $2.7\sigma$ to
   $1\sigma$; for the single scalar model we have used
   $m_{\Psi_\ell}=105\,\mathrm{GeV}$ and $m_{\Phi}=62\, \mathrm{GeV}$,
   while for the two scalar models we used
   $m_{\Psi_\ell}=105\,\mathrm{GeV},\,m_{\Phi_L}=62\, \mathrm{GeV},\,
   m_{\Phi_H}=400\,\mathrm{GeV}$.  The terms in parenthesis correspond
   to non-maximal mixing with $\theta=\pi /8$ in
   Eq.~(\ref{mixing}). \label{comparison1} }
\end{center}
\end{table}%

\subsection{Comparison}

In Table \ref{comparison1} we compare the results obtained above for
the benchmark points. The first of the two columns under ``Two
scalar'', $V^D_{2b}=0$, corresponds to the pure scalar induced mixing
case. For the  two scalar model we assume the bound from $K_L-K_S$
mass difference, $|\widetilde\Gamma_s|\le0.57$ is saturated, and we
show, in parenthesis, values  corresponding to non-maximal scalar mixing with $\theta=\pi /8$. 
In all cases we used $V^D_{2s}\approx1.0$, $V^D_{3d}\approx0$,
$V^D_{2d}\approx0.23$.  For the single scalar model we have used
$m_{\Psi_\ell} = m_{\Phi}  =100\,\mathrm{GeV}$, 
  and $m_{\Psi_Q} = 2.0\,\mathrm{TeV}$, while  for the two-scalar models we used  $m_{\Psi_\ell} = m_{\Phi_L}
= 100\,\mathrm{GeV}$, 
$m_{\Phi_H} = 400\,\mathrm{GeV}$ and $m_{\Psi_Q} =
2.0\,\mathrm{TeV}$. In the two-scalar models, $b\to s\mu\mu$ is computed at the benchmark point  $|\Gamma _b| =
1.5$. For $\Delta a_\mu$ the parameter values given reduce the
discrepancy in $(g-2)_\mu$ from $2.7\sigma$ to $1\sigma$, and, as
explained in Secs.~\ref{subSec:amu} and~\ref{DM}, for the
single scalar model we have used a different benchmark point, namely, $m_{\Psi_\ell}=105\,\mathrm{GeV}$ and
$m_{\Phi}=62\, \mathrm{GeV}$, while for the two scalar models we used
$m_{\Psi_\ell}=105\,\mathrm{GeV},\,m_{\Phi_L}=62\, \mathrm{GeV},\,
m_{\Phi_H}=400\,\mathrm{GeV}$.

\subsection{$\mu\rightarrow e\gamma$}\label{LFV}

Following \cite{Arnan:2016cpy}
the decay $\mu\to e\gamma$ is described by the effective Hamiltonian
\begin{equation}
{\cal H}_{\rm eff}^{\mu \to e\gamma}={-}C_{\mu \to e\gamma}  m_\mu 
 (\bar{e}\sigma^{\mu\nu} P_R \mu)\, F_{\mu \nu},
\end{equation}
from which the branching ratio is obtained according to
\bea
\textrm{Br}\left( {\mu  \to e\gamma } \right) = \frac{{m_\mu ^5}}{{4\pi }}{\tau _\mu }
|C_{\mu \to e\gamma}|^2,
\eea
where $\tau_\mu$ denotes the life-time of the muon. The experimental
upper limit~\chgdone{\cite{Adam:2013mnn}}{\cite{Adam:2013mnn,Cei:2017puj}}  is currently given by
\begin{equation}
 \textrm{Br}^{\textrm{exp}}\left( {\mu  \to e\gamma } \right) \leq 4.2\times 10^{-13}\chgdone{}{\,,\qquad\text{(90\% C.L.)}, } 
\end{equation}
giving the bound
\begin{equation}
   m_\mu^2|C_{\mu \to e\gamma}|\,<\,3.9\times 10^{-15}\,.
\end{equation}

The coefficient $C_{\mu \to e\gamma}$ is related 
to the  scalar contribution to the anomalous magnetic moment of the muon,
\begin{equation}
  C_{\mu \to e\gamma}\,\approx\,\frac{e}{m_\mu^2}V^E_{2e}\Delta a_\mu\,,
  \label{eq:relemu}
  \end{equation}
  where $V^E_{2e}$ is the component of the electron in the muon
  current eigenstate. Large enough $\Delta a_\mu$ to explain the
  discrepancy of the SM prediction with experiment requires
  $V^E_{2e}<10^{-5}$. Its value is model dependent. In the model of
  Sec.~\ref{family}, $V^E_{2e}\propto (\chi/M_q)^4$. Given the
  uncertainty in the string of couplings and messenger masses
  generating this term such a suppression may be sufficient to satisfy
  current experimental bounds. In the model presented in
  App.~\ref{app:noLepMix} the mixing vanishes.

\subsection{$Z\rightarrow\mu^+\mu^-$}
In the two scalar model the $Z$ penguin contribution  involving $\Psi_\ell$ and $\Phi$ \chgdone{gives}{to the correction to the coupling of the
  left handed muon to the $Z$-boson is given by \cite{Arnan:2016cpy}}
\begin{equation}
\frac{\delta g_{L\,\mu}}{g_{L\,\mu}^{\textrm{SM}}}(q^2)\;=\;-\frac{1}{64\pi^2}\frac{q^2}{m_{\Psi_\ell}^2}
     \chgdone{\Gamma _{\mu}^2}{|\Gamma _{\mu}|^2}\;\left ( \widetilde G_9(x_{\Phi_L} )+\widetilde G_9(x_{\Phi_H} )\right )
      \label{eq:Zmumu}
\end{equation}
where \chgdone{$g_{L\mu}$ is the LH $Z\mu\mu$ coupling and}{}
\begin{equation}
G_{9}^{}(x)=\dfrac{7 -36x+45x^2-16x^3 +6(2x-3)x^2\log x}{36(x-1)^4},\;\;\widetilde G_9(x)=x^{-1} G_9(x^{-1})\,.
\end{equation}
The LEP measurement \cite{ALEPH:2005ab} $\;g_{L\,\mu}^{\textrm{exp}}(m_Z^2)=-0.2689\pm 0.0011$ implies
\begin{equation}
  \left|\frac{\delta g_{L\,\mu}}{g_{L\,\mu}^{\textrm{SM}}}(m_Z^2)\right| \leq 0.8\%\;\;\;(2\sigma).
  \label{eq:Zmumu_LEP}
\end{equation}
For the case $m_{\Psi_\ell}=100\;\text{GeV}$,
$m_{\Phi_L}=80\;\text{GeV}$, $m_{\Phi_H}=400\;\text{GeV}$, we find
$\frac{\delta g_{L\,\mu}}{g_{L\,\mu}^{\textrm{SM}}}(M_Z^2)=9\times
10^{-4}|\Gamma_{\mu}|^2$, well within the LEP bound for perturbative
couplings. For the single scalar case the new contribution is smaller.

\section{Bounds on the new particle masses and their dark matter abundance}\label{DM}
In the Abelian family symmetry models  of Sec.~\ref{subsec:DM} only the heavy vector-like quark and lepton states and the scalar states $\Phi$ and $\tilde\Phi$ are odd under the $Z_2$ symmetry. As a result the lightest $Z_2$ odd state  will be stable and a candidate for dark matter.  In the mass range of interest here direct searches exclude the possibility that the DM candidate is the heavy lepton so one is left with the possibility that the lightest scalar is the DM.  

In this case the bounds on the heavy quark mass come from pair
production of the coloured vector-like quarks and their subsequent
decay into the light scalar plus a  quark ($pp \to \Psi_Q \bar
\Psi_Q$; $\Psi_Q \to q \Phi_L$) where $q$ is dominantly a third
generation quark. The LHC limits for this case have been studied in
detail   by Kawamura et al. \cite{Kawamura:2017ecz}\footnote{See also
  \cite{Chala:2017xgc}.} and their results are shown in their Fig.~1, case A. One sees that for a scalar mass of $O(100)\;$GeV the heavy quark should be heavier than $1.6\;$TeV. 

For the model with just a single scalar, $\Phi$, the possibility it is
dark matter has been studied in detail by  Calibbi et
al. \cite{Calibbi:2018rzv}. The model corresponds to their LL1 model
and in Fig.~4 they show the dark matter abundance in the $m_S\;(\equiv m_\Phi)$, $ m_L(\equiv m_{\Psi_\ell})$ plane for a range of Yukawa couplings, $\lambda_L\;(\equiv \Gamma_\mu)$. In Fig.~6 they plot the dark matter abundance in the $\lambda_L$, $m_S$ plane for a couple of values for $m_L$. Also shown are the LHC limits on $m_S$ and the $1\sigma$ and $2\sigma$ limits on $(g-2)_\mu$. 

One may see that there is a small window with $m_\Phi=(60-80) \;\text{GeV}$ which has not been excluded by the LHC.
Interestingly it is possible that $\Phi$ makes up all of dark matter in this mass range, depending on the heavy lepton mass and its coupling to the scalar. For example this is possible for $m_{\Psi_\ell}=105\;$GeV, a scalar mass $m_{\Phi_L}=62\;$GeV and $\Gamma_\mu=1.2$, a value a bit below the allowed range coming from $b\rightarrow s\mu\mu$. With this Yukawa coupling the discrepancy in the anomalous magnetic moment is reduced to $1.5\sigma$.

For the two scalar model the window applies to  the light scalar mass, $m_{\Phi_L}=(60-80)\;\text{GeV}$. In this case, for maximal mixing, $\lambda_L\equiv \Gamma_\mu/\sqrt{2}$ so, for a heavy lepton mass of $105$ GeV, 
the light scalar can make up all of dark matter for $\Gamma_\mu=1.7$. This is in the range required to explain 
$b\rightarrow s\mu\mu$ and within $1.3\sigma$ of the anomalous magnetic moment. The latter is quite sensitive to the mass of the heavier scalar and the discrepancy falls to $1\sigma$ for $m_{\Phi_H}=200\;\text{GeV}$.

Note that the bounds on the vector-like quarks and leptons and the new scalars are significantly weakened in the case discussed in App.~\ref{app:no-DM} when the lightest new state can decay rapidly into Standard Model states.

\section{Summary and Conclusions}\label{sum}
In this paper we have explored the possibility that the neutral current anomalies observed  in B decays come from loop effects generated by Yukawa couplings of SM fermions to new heavy SM singlet  scalars and new heavy vector-like quark and lepton SM doublets. We have shown that the phenomenologically required structure of the Yukawa couplings can be ensured by a simple Abelian family symmetry and that the same symmetry can  generate and acceptable pattern of fermion masses and mixing angles, thus relating the lepton non-universality observed in the B-sector to the non-universality of the fermion mass matrix.

We have analysed two simple models, one with a single heavy scalar in
which flavour changing is driven solely by the mixing in the fermion
sector and a second involving two heavy scalars in which flavour
changing is generated by  fermion mixing and/or scalar mixing. In both
cases, for the vector-like quark much heavier than the other new
states, it is possible to generate the effective operators, $O_9$ and
$O_{10}$ with coefficient large enough to  explain the B-anomaly
without violating the stringent bound coming from $B_s-\bar B_s$
mixing. Indeed in the two scalar model it is possible to avoid this
bound completely
with the caveat that one also has a strong cancellation in $K-\bar K$ and $D - \bar
D$ mixing.  We have also checked that the Yukawa couplings needed do not generate an unacceptable contribution to the  $K_L-K_S$ mass difference and $D-\bar{D}$ mixing or to other lepton family number or universality violating processes. While the Yukawa couplings required are quite large, they remain in the perturbative regime with the further new states needed to avoid their Landau poles only required abovel a scale of $O(10^6)\;$GeV. It is interesting to note that the same coupling generates a contribution to the muon anomalous magnetic moment that resolves the observed discrepancy of the experimental value with the SM prediction. For this to be the case the lightest scalar and vector-like lepton should be quite light, of $O(100)\;$GeV, but if they decay rapidly to SM states there will be no significant missing energy signal and the best limit is the LEP limit for the heavy lepton at $100\;$GeV. The scalar cannot be directly pair produced and so the limit on its mass is much milder.

On the other hand  a simple $Z_2$ extension of the family symmetry ensures that the new states are distinguished from the SM states so that they can only be produced in pairs and the lightest state, which must be a scalar, is stable and a DM candidate. In the case of the two scalar model the DM abundance for the Yukawa coupling needed to explain the B-anomalies is at the level that it can make up all of DM for a scalar mass of $O(100)\;$GeV. Of course it is necessary to check that the new heavy states should not have been observed to date and in this case there are significant missing energy signals. The bounds on the coloured quarks are quite severe, requiring them to be above $1.5\;$TeV and accordingly our numerical estimates were done for a $2\;$TeV vector-like doublet of quarks. The other states must be much lighter  and it turns out that there is a viable window with the lightest scalar mass of $O(60-80)\;$GeV and the heavy lepton mass of $O(100)\;$GeV. In this region it is possible simultaneously to explain the neutral B anomalies, the muon anomalous magnetic moment and the DM abundance. It should be possible for the LHC experimental searches to fully explore this region in the near future.

\vspace{1cm}
 
Acknowledments: We are grateful to Alan Barr, Andrzej Buras, Yosi Nir, Gilad
Perez, Andreas Weiler, Robert Ziegler and especialy Andreas Crivellin and Kazuki
Sakurai for helpful discussions and to the Theory Group, CERN, where much of the work was done..  The work of SP is partially
supported by the National Science Centre, Poland, under research
grants DEC-2015/18/M/ST2/00054, DEC-2015/19/B/ST2/02848 and
DEC-2016/23/G/ST2/04301. The work of BG is partially supported by the
Department of Energy, USA, under research grant DE-SC0009919.

\vspace{1cm}

\appendix{\huge \bf Appendix}
\vspace{0.5cm}

\section{Elimination of lepton flavour changing processes}
\label{app:noLepMix}

While the lepton family mixing is small enough to avoid violating  bounds on  lepton flavour violating processes, $\tau\rightarrow \mu \gamma$, $\mu\rightarrow e \gamma$ etc. the expectation in this model is that the branching ratios are close to these  bounds. Thus it is of interest to check whether these can be further suppressed. Indeed this is relatively simple through the introduction of a further family symmetry. A simple example is to add a $Z_3$ discrete symmetry under which the 3 different lepton families transform as $1,\;\alpha$ and $\alpha^2$ where $\alpha$ is the cube root of unity. In this case there can be no lepton family mixing and the current eigenstates are the mass eigenstates. If, in addition,  $\Psi_{\ell,(L,R)}$ transforms in the same way as the muon under $Z_3$ then the box graph will only involve muons and generate $b\rightarrow s\mu\mu$ as desired. Because the lepton families do not mix it is possible to simplify their charge structure under $U(1)_F$. The choice given in Eq.~(\ref{Z3model}) gives an acceptable lepton mass hierarchy with ${m_\mu\over m_\tau}=O(\epsilon)$ and ${m_e\over m_\mu}=O(\epsilon^3)$.
\be
\begin{array}{*{20}{c}}
{}&\vline& Q_1&q_1&{{Q_2}}&{{q_2}}&{{Q_3}}&{{q_3}}&L_1&l_1&{{L_2}}&{{l_2}}&{{L_3}}&{{l_3}}&\Psi_{Q,(L,R)}&\Psi_{\ell,(L,R)}&\Phi&\tilde\Phi &H&\chi\\
\hline
Q_F&\vline& 3&-1&2&0&0&0&2&-2&-1&0&0&0&2&1&-2&0&0&1\\
\hline 
Z_3&\vline&1(\alpha)&1(\alpha)&1&1&1&1&1&1&\alpha&\alpha&\alpha^2&\alpha^2&1&\alpha&1&1&1&1
\end{array}
\label{Z3model}
\ee

With these $Z_3$ charges given in Eq.~(\ref{Z3model}) there is no lepton family mixing. The muon mass remains of $O(\epsilon)$ relative to the $\tau$ mass but the electron  mass  is now $O(\epsilon^3)$ relative to the muon mass, still phenomenologically acceptable given the unknown $O(1)$ couplings involved. 

The quark masses and mixings are unchanged compared to the model discussed in Sec.~\ref{family} with couplings of $Q_1$ to the vector-like quark generating additional non-universal lepton decay modes.  These can readily be forbidden if the first generation quark doublet and down quark transform non-trivially under $Z_3$  as shown by the terms in brackets in Eq.~(\ref{Z3model}). In this case the mixing to the first generation of quarks must occur in the up quark sector. 

Neutrino massese are presumed to be generated at very short
distances. The $Z_3$ symmetry only allows $L_2L_3$ terms in a  Weinberg operator
for see-saw neutrino masses. However, it  is natural to expect that
the dynamics responsible for breaking lepton number does not respect
$Z_3$. 

\vspace{0.5cm}

\section{Simplified model without a dark matter candidate} 
\label{app:no-DM}

 If one is not concerned to have the possibility that the new heavy
 states are dark matter candidates it is possible to simplify the
 original model by dropping the $Z_2$ symmetry. In this case there are
 additional terms allowed, for example
\begin{multline}
  {\cal L}_{eff}^M=
  M'\overline Q_2 P_R\Psi_{Q} +\overline \Psi_{Q}P_R \;q_{2}\; H\;({\chi\over M_q})^2
  + \overline \Psi_{Q} P_R\;q_{3}\; H\;({\chi\over M_q})^2
  + \overline{\Psi}_{Q}P_R \;q_{1}\; H\;({\chi\over M_q})^3
  +\widetilde\Gamma_s\overline \Psi_QP_R\Psi_Q\tilde\Phi
\label{oops}
\end{multline}
The first four terms induce mixing between $q_i$, $Q_2$ and $\Psi_Q$;
together with the corresponding mass terms in Sec.~\ref{family} these
induce $4\times4$ mass matrices for up and down type quarks so that,
{\it e.g.}, the unitary matrices that diagonalize these are $4\times4$. The
last term is an interaction, and both this term and the mass mixing
clearly allow the original $Z_2$ odd
states to decay so they are no longer dark matter candidates.


\begin{thebibliography}{99}
\bibitem{Aaij:2014ora}
  R.~Aaij {\it et al.} [LHCb Collaboration],
  Phys.\ Rev.\ Lett.\  {\bf 113} (2014) 151601
  doi:10.1103/PhysRevLett.113.151601
  [arXiv:1406.6482 [hep-ex]].
\bibitem{Aaij:2017vbb}
  R.~Aaij {\it et al.} [LHCb Collaboration],
  JHEP {\bf 1708} (2017) 055
  doi:10.1007/JHEP08(2017)055
  [arXiv:1705.05802 [hep-ex]].
\bibitem{Aaij:2013qta}
  R.~Aaij {\it et al.} [LHCb Collaboration],
  Phys.\ Rev.\ Lett.\  {\bf 111} (2013) 191801
  doi:10.1103/PhysRevLett.111.191801
  [arXiv:1308.1707 [hep-ex]].
\bibitem{Aaij:2015oid}
  R.~Aaij {\it et al.} [LHCb Collaboration],
  JHEP {\bf 1602} (2016) 104
  doi:10.1007/JHEP02(2016)104
  [arXiv:1512.04442 [hep-ex]].
\bibitem{Wehle:2016yoi}
  S.~Wehle {\it et al.} [Belle Collaboration],
  Phys.\ Rev.\ Lett.\  {\bf 118} (2017) no.11,  111801
  doi:10.1103/PhysRevLett.118.111801
  [arXiv:1612.05014 [hep-ex]].
\bibitem{Aaij:2014pli}
  R.~Aaij {\it et al.} [LHCb Collaboration],
  JHEP {\bf 1406} (2014) 133
  doi:10.1007/JHEP06(2014)133
  [arXiv:1403.8044 [hep-ex]].
\bibitem{Aaij:2015esa}
  R.~Aaij {\it et al.} [LHCb Collaboration],
  JHEP {\bf 1509} (2015) 179
  doi:10.1007/JHEP09(2015)179
  [arXiv:1506.08777 [hep-ex]].
\bibitem{Altmannshofer:2017yso}
  W.~Altmannshofer, P.~Stangl and D.~M.~Straub,
  Phys.\ Rev.\ D {\bf 96} (2017) no.5,  055008
  doi:10.1103/PhysRevD.96.055008
  [arXiv:1704.05435 [hep-ph]].
\bibitem{DAmico:2017mtc}
  G.~D'Amico, M.~Nardecchia, P.~Panci, F.~Sannino, A.~Strumia, R.~Torre and A.~Urbano,
  JHEP {\bf 1709} (2017) 010
  doi:10.1007/JHEP09(2017)010
  [arXiv:1704.05438 [hep-ph]].
\bibitem{Hiller:2017bzc}
  G.~Hiller and I.~Nisandzic,
  Phys.\ Rev.\ D {\bf 96} (2017) no.3,  035003
  doi:10.1103/PhysRevD.96.035003
  [arXiv:1704.05444 [hep-ph]].

\bibitem{Capdevila:2017bsm}
  B.~Capdevila, A.~Crivellin, S.~Descotes-Genon, J.~Matias and J.~Virto,
  JHEP {\bf 1801} (2018) 093
  doi:10.1007/JHEP01(2018)093
  [arXiv:1704.05340 [hep-ph]].


\bibitem{Geng:2017svp}
  L.~S.~Geng, B.~Grinstein, S.~Jäger, J.~Martin Camalich, X.~L.~Ren and R.~X.~Shi,
  Phys.\ Rev.\ D {\bf 96} (2017) no.9,  093006
  doi:10.1103/PhysRevD.96.093006
  [arXiv:1704.05446 [hep-ph]].
\bibitem{Ciuchini:2017mik}
  M.~Ciuchini, A.~M.~Coutinho, M.~Fedele, E.~Franco, A.~Paul, L.~Silvestrini and M.~Valli,
  Eur.\ Phys.\ J.\ C {\bf 77} (2017) no.10,  688
  doi:10.1140/epjc/s10052-017-5270-2
  [arXiv:1704.05447 [hep-ph]].

\bibitem{Froggatt:1978nt}
  C.~D.~Froggatt and H.~B.~Nielsen,
  Nucl.\ Phys.\ B {\bf 147} (1979) 277.

 \bibitem{Gripaios:2015gra}   B.~Gripaios, M.~Nardecchia and S.~A.~Renner,   
  JHEP {\bf 1606} (2016) 083
  doi:10.1007/JHEP06(2016)083
  [arXiv:1509.05020 [hep-ph]].
 
  \bibitem{Cline:2017qqu}   J.~M.~Cline and J.~M.~Cornell,   
  Phys.\ Lett.\ B {\bf 782} (2018) 232
  doi:10.1016/j.physletb.2018.05.034
  [arXiv:1711.10770 [hep-ph]].
  
  \bibitem{Poh:2017tfo}   Z.~Poh and S.~Raby,   
  Phys.\ Rev.\ D {\bf 96} (2017) no.1,  015032
  doi:10.1103/PhysRevD.96.015032
  [arXiv:1705.07007 [hep-ph]].
  
   \bibitem{Dhargyal:2017vcu}   L.~Dhargyal,   
  arXiv:1711.09772 [hep-ph].
   \bibitem{Dhargyal:2018bbc}   L.~Dhargyal and S.~K.~Rai,   
  arXiv:1806.01178 [hep-ph].




\bibitem{Arnan:2016cpy}
  P.~Arnan, L.~Hofer, F.~Mescia and A.~Crivellin,
  JHEP {\bf 1704} (2017) 043
  doi:10.1007/JHEP04(2017)043
  [arXiv:1608.07832 [hep-ph]].

\bibitem{Das:2017kfo} 
  D.~Das, C.~Hati, G.~Kumar and N.~Mahajan,
  Phys.\ Rev.\ D {\bf 96}, no. 9, 095033 (2017)
  doi:10.1103/PhysRevD.96.095033
  [arXiv:1705.09188 [hep-ph]].


\bibitem{Lees:2012xj}
  J.~P.~Lees {\it et al.} [BaBar Collaboration],
  Phys.\ Rev.\ Lett.\  {\bf 109} (2012) 101802
  doi:10.1103/PhysRevLett.109.101802
  [arXiv:1205.5442 [hep-ex]].

\bibitem{Lees:2013uzd}
  J.~P.~Lees {\it et al.} [BaBar Collaboration],
  Phys.\ Rev.\ D {\bf 88} (2013) no.7,  072012
  doi:10.1103/PhysRevD.88.072012
  [arXiv:1303.0571 [hep-ex]].

\bibitem{Huschle:2015rga}
  M.~Huschle {\it et al.} [Belle Collaboration],
  Phys.\ Rev.\ D {\bf 92} (2015) no.7,  072014
  doi:10.1103/PhysRevD.92.072014
  [arXiv:1507.03233 [hep-ex]].

\bibitem{Sato:2016svk}
  Y.~Sato {\it et al.} [Belle Collaboration],
  Phys.\ Rev.\ D {\bf 94} (2016) no.7,  072007
  doi:10.1103/PhysRevD.94.072007
  [arXiv:1607.07923 [hep-ex]].

\bibitem{Hirose:2016wfn}
  S.~Hirose {\it et al.} [Belle Collaboration],
  Phys.\ Rev.\ Lett.\  {\bf 118} (2017) no.21,  211801
  doi:10.1103/PhysRevLett.118.211801
  [arXiv:1612.00529 [hep-ex]].

\bibitem{Hirose:2017dxl}
  S.~Hirose {\it et al.} [Belle Collaboration],
  Phys.\ Rev.\ D {\bf 97} (2018) no.1,  012004
  doi:10.1103/PhysRevD.97.012004
  [arXiv:1709.00129 [hep-ex]].

\bibitem{Aaij:2015yra}
  R.~Aaij {\it et al.} [LHCb Collaboration],
  Phys.\ Rev.\ Lett.\  {\bf 115} (2015) no.11,  111803
   Erratum: [Phys.\ Rev.\ Lett.\  {\bf 115} (2015) no.15,  159901]
  doi:10.1103/PhysRevLett.115.159901, 10.1103/PhysRevLett.115.111803
  [arXiv:1506.08614 [hep-ex]].

\bibitem{Aaij:2017uff}
  R.~Aaij {\it et al.} [LHCb Collaboration],
  Phys.\ Rev.\ Lett.\  {\bf 120} (2018) no.17,  171802
  doi:10.1103/PhysRevLett.120.171802
  [arXiv:1708.08856 [hep-ex]].

\bibitem{Aaij:2017deq}
  R.~Aaij {\it et al.} [LHCb Collaboration],
  Phys.\ Rev.\ D {\bf 97} (2018) no.7,  072013
  doi:10.1103/PhysRevD.97.072013
  [arXiv:1711.02505 [hep-ex]].

\bibitem{Aaij:2017tyk}
  R.~Aaij {\it et al.} [LHCb Collaboration],
  Phys.\ Rev.\ Lett.\  {\bf 120} (2018) no.12,  121801
  doi:10.1103/PhysRevLett.120.121801
  [arXiv:1711.05623 [hep-ex]].

\bibitem{Alonso:2015sja}
  R.~Alonso, B.~Grinstein and J.~Martin Camalich,
  JHEP {\bf 1510} (2015) 184
  doi:10.1007/JHEP10(2015)184
  [arXiv:1505.05164 [hep-ph]].

\bibitem{Hiller:2014yaa}
  G.~Hiller and M.~Schmaltz,
  Phys.\ Rev.\ D {\bf 90} (2014) 054014
  doi:10.1103/PhysRevD.90.054014
  [arXiv:1408.1627 [hep-ph]].

  
\bibitem{Barbieri:2015yvd}
  R.~Barbieri, G.~Isidori, A.~Pattori and F.~Senia,
  Eur.\ Phys.\ J.\ C {\bf 76} (2016) no.2,  67
  doi:10.1140/epjc/s10052-016-3905-3
  [arXiv:1512.01560 [hep-ph]].


\bibitem{Assad:2017iib}
  N.~Assad, B.~Fornal and B.~Grinstein,
  Phys.\ Lett.\ B {\bf 777} (2018) 324
  doi:10.1016/j.physletb.2017.12.042
  [arXiv:1708.06350 [hep-ph]].

\bibitem{DiLuzio:2017vat}
  L.~Di Luzio, A.~Greljo and M.~Nardecchia,
  Phys.\ Rev.\ D {\bf 96} (2017) no.11,  115011
  doi:10.1103/PhysRevD.96.115011
  [arXiv:1708.08450 [hep-ph]].

\bibitem{Calibbi:2017qbu}
  L.~Calibbi, A.~Crivellin and T.~Li,
  arXiv:1709.00692 [hep-ph].

\bibitem{Barbieri:2017tuq}
  R.~Barbieri and A.~Tesi,
  Eur.\ Phys.\ J.\ C {\bf 78} (2018) no.3,  193
  doi:10.1140/epjc/s10052-018-5680-9
  [arXiv:1712.06844 [hep-ph]].

\bibitem{Blanke:2018sro}
  M.~Blanke and A.~Crivellin,
  Phys.\ Rev.\ Lett.\  {\bf 121} (2018) no.1,  011801
  doi:10.1103/PhysRevLett.121.011801
  [arXiv:1801.07256 [hep-ph]].

\bibitem{ACrivellin:2018qbd}
A.~Crivellin, Ch.~Greub, F.~Saturnino and D.~Mueller,
 arXiv:1807.02068 [hep-ph].

\bibitem{Greljo:2018ogz}
  A.~Greljo, D.~J.~Robinson, B.~Shakya and J.~Zupan,
  JHEP {\bf 1809} (2018) 169
  doi:10.1007/JHEP09(2018)169
  [arXiv:1804.04642 [hep-ph]].


 
\bibitem{Amhis:2016xyh}
  Y.~Amhis {\it et al.} [HFLAV Collaboration],
  Eur.\ Phys.\ J.\ C {\bf 77} (2017) no.12,  895
  doi:10.1140/epjc/s10052-017-5058-4
  [arXiv:1612.07233 [hep-ex]].

\bibitem{Adam:2013mnn}
  J.~Adam {\it et al.} [MEG Collaboration],
  Phys.\ Rev.\ Lett.\  {\bf 110} (2013) 201801
  doi:10.1103/PhysRevLett.110.201801
  [arXiv:1303.0754 [hep-ex]].

\bibitem{Cei:2017puj}
  F.~Cei {\it et al.} [MEG Collaboration],
  PoS NEUTEL {\bf 2017} (2018) 023.
  doi:10.22323/1.307.0023

  
\bibitem{Chatrchyan:2013oca}
  S.~Chatrchyan {\it et al.} [CMS Collaboration],
  JHEP {\bf 1307} (2013) 122
  doi:10.1007/JHEP07(2013)122
  [arXiv:1305.0491 [hep-ex]].  
  
\bibitem{Aad:2015eda}
  G.~Aad {\it et al.} [ATLAS Collaboration],
  Phys.\ Rev.\ D {\bf 93} (2016) no.5,  052002
  doi:10.1103/PhysRevD.93.052002
  [arXiv:1509.07152 [hep-ex]].
  
\bibitem{Aaboud:2017vwy}
  M.~Aaboud {\it et al.} [ATLAS Collaboration],
  arXiv:1712.02332 [hep-ex].
  
  
  
  
  
  
   
\bibitem{ALEPH:2005ab}S.~Schael {\it et al.} [ALEPH and DELPHI and L3 and OPAL and SLD Collaborations and LEP Electroweak Working Group and SLD Electroweak Group and SLD Heavy Flavour Group],
Phys.\ Rept.\  {\bf 427} (2006) 257doi:10.1016/j.physrep.2005.12.006[hep-ex/0509008].
  
  
  \bibitem{Chala:2017xgc}   M.~Chala,   
  Phys.\ Rev.\ D {\bf 96} (2017) no.1,  015028
  doi:10.1103/PhysRevD.96.015028
  [arXiv:1705.03013 [hep-ph]].
  
\bibitem{Kawamura:2017ecz}
  J.~Kawamura, S.~Okawa and Y.~Omura,
  Phys.\ Rev.\ D {\bf 96} (2017) no.7,  075041
  doi:10.1103/PhysRevD.96.075041
  [arXiv:1706.04344 [hep-ph]].

\bibitem{Kowalska:2017iqv}
  K.~Kowalska and E.~M.~Sessolo,
  JHEP {\bf 1709} (2017) 112
  doi:10.1007/JHEP09(2017)112
  [arXiv:1707.00753 [hep-ph]].

\bibitem{Calibbi:2018rzv}
  L.~Calibbi, R.~Ziegler and J.~Zupan,
  JHEP {\bf 1807} (2018) 046
  doi:10.1007/JHEP07(2018)046
  [arXiv:1804.00009 [hep-ph]].
  
   \end{thebibliography}
     \end{document}